\documentclass[12pt]{article}

\usepackage[margin=1in]{geometry}

\usepackage{libertine}
\usepackage[libertine]{newtxmath}
\usepackage[scaled=0.85]{inconsolata}
\usepackage[T1]{fontenc}

\usepackage{amsmath}
\usepackage{booktabs}
\usepackage{tikz}
\usepackage{comment}
\usepackage{natbib}
\usepackage{hyperref}
\usepackage{xcolor}
\usepackage{microtype}
\usepackage{parskip}
\usepackage{fancyvrb}
\usetikzlibrary{positioning, arrows.meta, shapes, calc}

\hypersetup{
  colorlinks=true,
  linkcolor=blue!60!black,
  citecolor=blue!60!black,
  urlcolor=blue!60!black
}

\newcommand{\pkg}[1]{\textbf{#1}}
\newcommand{\proglang}[1]{\textsf{#1}}
\newcommand{\code}[1]{\texttt{#1}}

\definecolor{outputgray}{RGB}{100,100,100}

\newenvironment{CodeChunk}{%
  \vspace{4pt}%
}{%
  \vspace{4pt}%
}

\DefineVerbatimEnvironment{CodeInput}{Verbatim}{%
  fontsize=\small,
  fontfamily=tt,
  frame=leftline,
  framerule=2pt,
  rulecolor=\color{black!50},
  framesep=4pt,
  xleftmargin=6pt
}

\DefineVerbatimEnvironment{CodeOutput}{Verbatim}{%
  fontsize=\small,
  fontfamily=tt,
  formatcom=\color{outputgray},
  frame=leftline,
  framerule=2pt,
  rulecolor=\color{gray!40},
  framesep=4pt,
  xleftmargin=6pt
}

\title{%
  \textsf{PyINLA}: Fast Bayesian Inference for Latent\\
  Gaussian Models in \textsf{Python}
}

\author{
  Esmail Abdul Fattah\thanks{Corresponding author:
  \texttt{esmail.abdulfattah@kaust.edu.sa}},\quad
  Elias Krainski,\quad
  H{\aa}vard Rue \\[0.3em]
  \small CEMSE Division, King Abdullah University of 
  Science and Technology (KAUST), \\
  \small Thuwal, 23955, Makkah, Saudi Arabia
}

\date{}

\begin{document}
\maketitle

\begin{abstract}
\noindent
Bayesian inference often relies on Markov chain Monte Carlo (MCMC) methods, particularly required for non-Gaussian data families.
When dealing with complex hierarchical models, the MCMC approach can be computationally demanding in workflows that require repeated model fitting or when working with models of large dimensions with limited hardware resources. 
The Integrated Nested Laplace Approximations (INLA) is a deterministic alternative for models with non-Gaussian data that belong to the class of latent Gaussian models (LGMs), yielding accurate approximations to posterior marginals in many applied settings. 
The INLA method was implemented in \proglang{C} as a standalone program, \textbf{inla}, that is 
widely used in \proglang{R} through the \pkg{INLA} package. 
This paper introduces \textsf{PyINLA}, a dedicated \proglang{Python} package that provides a Pythonic interface directly to the \textbf{inla} program. 
Therefore, \textsf{PyINLA} enables specifying LGMs, running INLA-based inference, and accessing posterior summaries directly from \proglang{Python} while leveraging the established INLA implementation. 
We describe the package design and illustrate its use on representative models, including generalized linear mixed models, time series forecasting, disease mapping, and geostatistical prediction, demonstrating how deterministic Bayesian inference can be performed in Python using INLA in a way that integrates naturally with common scientific computing workflows.
\end{abstract}


\section{Introduction}\label{sec:introduction}

Bayesian hierarchical models are widely used in applied statistical inference, offering a coherent framework for accounting for prior information and quantifying uncertainty. In many application areas, including spatial epidemiology \citep{Lawson2018}, environmental and ecological modeling \citep{Cressie2011}, and econometrics \citep{Koop2003}, the practical value of Bayesian modeling depends on computational tools that are fast, reliable, and easy to integrate into end-to-end analysis pipelines.

In contemporary software ecosystems, general-purpose Bayesian inference is most often performed using Markov chain Monte Carlo (MCMC) or variational inference (VI). These approaches underpin mature probabilistic programming frameworks, but they carry well-known trade-offs: MCMC is computationally intensive for models with high-dimensional latent structure and requires careful convergence assessment, while mean-field VI introduces approximation bias and tends to underestimate posterior uncertainty \citep{Brooks2011, Blei2017}. These trade-offs become acute in practice for hierarchical models with large latent fields, iterative workflows involving repeated model fitting (sensitivity analysis, model selection, cross-validation), and production settings that demand deterministic, reproducible outputs.

The Integrated Nested Laplace Approximations (INLA) provides a deterministic alternative for an important and practically broad class of models: the latent Gaussian models (LGMs) \citep{Rue2009}. The Laplace approximation exploits the conditional structure in the prior distribution of the latent field twice. One while searching for the posterior mode of the hyperparameters and then several nested Laplace approximations combined with numerical integration over the hyperparameters. 
These nested Laplace approximations in the INLA algorithm can be replaced by an implicit low-rank correction, 
as proposed in \cite{van2024vbc}, which is shown to be faster than the original INLA and as accurate, \cite{van2023new}.
When the latent precision matrix is sparse, as it is for Gaussian Markov random fields \citep{rue2005gaussian} and a wide class of structured random effects, INLA computes accurate approximations to posterior marginal distributions efficiently, even when the latent dimension is large. These properties have established INLA as a competitive method for spatial and spatiotemporal hierarchical models across applied statistics \citep{Blangiardo2013, Martins2013,competitionSpatial}.

The reference implementation of INLA is the \pkg{R-INLA} project
(\url{https://www.r-inla.org}), which has matured into a comprehensive
and widely used framework in the \proglang{R} ecosystem \citep{lindgren2015bayesian, bivand2015spatial, van2021new}. Several \proglang{R} packages extend its scope to broader spatial models, non-separable spatio-temporal structures, non-linear predictors, and joint longitudinal-survival models \citep{rSPDE, krainski2025inlaspacetime, bachl2019inlabru, lindgren2024inlabru, rustand2024joint}, with applications spanning biostatistics, geosciences, epidemiology, and environmental modeling \citep{alvares2024bayesian, fioravanti2023interpolating, myer2019spatiotemporal, seaton2024spatio}.

Practitioners in \proglang{Python}, now the dominant language in data science 
and machine learning, which need fast Bayesian inference for hierarchical models 
with spatial, temporal, or grouped structure have largely been limited to 
MCMC-based tools such as \pkg{PyMC} or \pkg{Stan}, which can be 
computationally demanding for models with large latent fields and requires 
careful convergence assessment. 
INLA provides a deterministic alternative 
for these models, but has until now lacked a native \proglang{Python} interface. 
\textsf{PyINLA} fills this gap: it is a self-contained \proglang{Python} package, 
with no \proglang{R} installation required, that brings INLA-based inference 
directly form the standalone \textbf{inla} \proglang{C} program into standard \proglang{Python} workflows built around \pkg{pandas}, 
\pkg{NumPy}, and \pkg{SciPy}.

This paper introduces \textsf{PyINLA}, a native \proglang{Python} package that brings INLA-based inference directly into the \proglang{Python} ecosystem. 
\textsf{PyINLA} exposes the compiled INLA computational engine through a Pythonic API, enabling users to specify LGMs, run inference, and access full posterior summaries using standard \proglang{Python} data structures (\pkg{pandas} DataFrames, \pkg{NumPy} arrays, \pkg{SciPy} sparse matrices). 
The package eliminates the need for an \proglang{R} installation or cross-language interoperability layers, making INLA accessible to the large and growing community of practitioners who build their analysis
and deployment pipelines in \proglang{Python}.

The main contributions of this work are:
\begin{itemize}
    \item \textbf{Native \proglang{Python} interface.} \textsf{PyINLA} provides a concise API for specifying LGMs using standard \proglang{Python} data
    structures, with results returned as \pkg{pandas} DataFrames and \pkg{NumPy} arrays ready for downstream analysis. No \proglang{R} installation or interoperability layer is required; the compiled \textbf{inla} engine is managed transparently by the package.
  \item \textbf{Broad LGM coverage.} The package currently supports widely used INLA likelihood families and latent model structures, including group-level random effects,
  temporal (random walks, AR processes), 
  and spatial (BYM, SPDE), 
  using principled prior specification via penalized complexity priors \citep{simpson2017penalising}.
 \item \textbf{Deterministic, reproducible inference.} 
 \textsf{PyINLA} returns deterministic approximations to posterior marginal distributions; thus no
 Monte Carlo variability, no convergence diagnostics requirement, and no need for stochastic seeds.
  \item \textbf{Seamless ecosystem integration.} Results are returned as structured \proglang{Python} objects that plug directly into \pkg{pandas}, \pkg{matplotlib}, \pkg{scikit-learn}, and other standard tools for downstream analysis, visualization, and deployment.
\end{itemize}

The broad LGM coverage is achieved by focusing on core LGM functionality, including support for the most popular likelihood families and latent models. Additional model components are planned for future releases. The current state of the art of the core numerical method is supported.

The remainder of this paper is organized as follows. Section~\ref{sec:background} reviews Bayesian computation in \proglang{Python} and positions \textsf{PyINLA} relative to commonly used alternatives. 
Section~\ref{sec:inla} summarizes the latent Gaussian model framework and the INLA approximation strategy, 
which suffice for understanding the package scope and outputs. Section~\ref{sec:pyinla} describes the design, implementation, and usage of \textsf{PyINLA}. Representative examples are provided in Section~\ref{sec:examples}, and Section~\ref{sec:conclusion} discusses limitations and future directions.

\section{Background: Bayesian computation and related software}\label{sec:background}

Applied Bayesian workflows involve iterative model development, repeated fitting across related datasets, and routine model checking. In these settings, the choice of inference method is driven not only by statistical objectives but also by runtime, diagnostic burden, and integration into data-processing and deployment pipelines.

Within \proglang{Python}, general-purpose Bayesian inference is commonly carried out using MCMC or VI, implemented in frameworks such as \pkg{PyMC}, \pkg{Stan} (via \pkg{CmdStanPy}), \pkg{TensorFlow Probability}, and \pkg{Pyro} \citep{Salvatier2016, Carpenter2017, Dillon2017, Bingham2019}. Modern MCMC methods such as Hamiltonian Monte Carlo and the No-U-Turn Sampler \citep{Hoffman2014} provide asymptotically exact inference but are expensive for hierarchical models with large latent structure and require substantial effort for tuning and convergence assessment. Variational methods replace sampling with optimization and reduce runtime, but they introduce bias and understate posterior uncertainty when posterior dependence is strong \citep{Rue2009,Zhang2019}. These trade-offs motivate deterministic alternatives for model classes where such methods are applicable.

INLA provides a deterministic alternative for latent Gaussian models by replacing sampling with analytical approximations, as outlined in Section~\ref{sec:introduction} and detailed in Section~\ref{sec:inla} \citep{Rue2009, Rue2017}. Empirical comparisons consistently demonstrate that INLA achieves comparable accuracy to MCMC while being one to two orders of magnitude faster across various applications, including disease mapping, clinical trials, and financial modeling \citep{de2015comparing, chen2024comparison, darkwah2022bayesian, nacinben2024multivariate}.

\textsf{PyINLA} provides a \proglang{Python}-native interface to the compiled INLA engine, making deterministic Bayesian inference for LGMs directly accessible within \proglang{Python} workflows.

\subsection{Positioning PyINLA relative to existing Python tools}

Table~\ref{tab:comparison} contrasts \textsf{PyINLA} with widely used general-purpose Bayesian tools in \proglang{Python}.

\begin{table}[t]
\centering
\caption{Comparison of Bayesian inference tools available in \proglang{Python}.}
\label{tab:comparison}
\small
\begin{tabular}{lcccc}
\toprule
 & \textsf{PyINLA} & \pkg{PyMC} & \pkg{Stan} & \pkg{TFP} \\
\midrule
Primary inference & INLA & MCMC/VI & MCMC/VI & MCMC/VI \\
Model class & LGM only & General & General & General \\
Monte Carlo error & No & Yes & Yes & Yes \\
Convergence diagnostics & Not needed & Required & Required & Required \\
Spatial models (SPDE) & Native & Manual & Manual & Manual \\
Deep learning integration & No & Indirect\textsuperscript{a} & No & Native \\
\bottomrule
\end{tabular}

\textsuperscript{a}Deep learning integration is possible via the PyTensor backend, but is not a primary design goal.
\end{table}

\pkg{PyMC} provides a flexible probabilistic programming interface with NUTS-based MCMC and VI. \pkg{Stan} offers a domain-specific language with state-of-the-art HMC/NUTS implementations. \pkg{TensorFlow Probability} integrates probabilistic modeling with the TensorFlow ecosystem, supporting workflows that combine Bayesian inference with deep learning. These tools offer broad modeling flexibility, but that generality comes at a cost: users must manage sampling efficiency, assess convergence, and absorb significant runtime for models with a complex hierarchical structure.

\textsf{PyINLA} takes a fundamentally different approach. By restricting 
attention to LGMs, it exploits the nested Laplace approximation combined 
with numerical methods for sparse matrices to compute 
posterior marginals without any sampling. 
For the large class of models within its scope, this design yields concrete practical advantages:
\begin{itemize}
  \item \textbf{Efficiency for structured latent fields.} Models with high-dimensional latent components (e.g., spatial and spatiotemporal effects) are fitted efficiently by exploiting sparsity in precision matrices.
  \item \textbf{Built-in structured components.} Common temporal and spatial components (e.g., random walks, autoregressive processes, CAR/BYM-type models, SPDE-based fields) are available with principled default priors.
\end{itemize}

The trade-off is scope: \textsf{PyINLA} is restricted to the latent Gaussian model class. Models with non-Gaussian latent structure, or likelihoods outside INLA support, require general-purpose tools. Applications that need direct samples from the full joint posterior may also favor MCMC, although \textsf{PyINLA} provides posterior sampling functionality for derived-quantity inference.

\section{The INLA methodology}\label{sec:inla}

This section summarizes the latent Gaussian model framework and the 
INLA approximation strategy at the level required to understand what 
\textsf{PyINLA} can represent and what it returns. 
Detailed methodological treatments are available in
\citet{Rue2009, abdul2023approximate, van2024vbc}.

Before the formal definition, it helps to build up the model 
structure from data. Suppose we observe data $y_i$ that could be
continuous measurements (e.g., temperature), counts (e.g., number 
of cases), or binary outcomes (e.g., success/failure). 
We want to model the expected response $\mu_i = \mathbb{E}[y_i]$ 
as a function of covariates and random effects. Since $\mu_i$ is 
often restricted (e.g., positive for counts, in $(0,1)$ for 
probabilities), a \emph{link function} $g(\cdot)$ maps it to the 
real line. The transformed response $\eta_i = g(\mu_i)$ is called 
the \emph{linear predictor}, modeled as:
\begin{equation}\label{eq:general-pattern-preview}
  \underbrace{\eta_i = g(\mu_i)}_{\text{linear predictor}}
  = \underbrace{\beta_0 + \beta_1 x_{1i} + \cdots}_{\text{fixed effects}}
  + \underbrace{u^{(1)}_{k_1(i)} + u^{(2)}_{k_2(i)} + \cdots}_{\text{random effects}},
  \qquad
  \underbrace{y_i \sim p(\cdot \mid \mu_i, \boldsymbol{\theta})}_{\text{likelihood}}.
\end{equation}
Here $\boldsymbol{\beta}$ are fixed regression coefficients 
and $u^{(k)}$ are structured random effects capturing variation 
due to spatial proximity, temporal trends, or group membership. 
Together, they form the \emph{latent field} 
$\boldsymbol{x} = (\beta_0, \boldsymbol{\beta}, 
\boldsymbol{u}_1, \ldots, \boldsymbol{u}_K)^\top$, 
which is assigned a joint Gaussian prior as detailed below.

\subsection{Latent Gaussian models}

A latent Gaussian model (LGM) can be defined following a hierarchical 
structure, which includes 1) a family (or likelihood) assumed for the 
data, 2) Gaussian priors for the parameters in the linear predictor 
(the latent field), and 3) a hyperparameter vector. 
Let $\boldsymbol{y}=(y_1,\ldots,y_n)^\top$ denote observed data, 
$\boldsymbol{x}\in\mathbb{R}^N$ the latent field as introduced 
above, and $\boldsymbol{\theta}\in\mathbb{R}^p$ hyperparameters.
A simplified view follows; see \citet{Rue2017} for full details.

\textbf{Observation model.} The data are conditionally independent 
given $\boldsymbol{x}$ and $\boldsymbol{\theta}$:
\begin{equation}
y_i \mid \boldsymbol{x},\boldsymbol{\theta} 
\stackrel{\mathrm{ind}}{\sim} 
p(y_i \mid \eta_i,\boldsymbol{\theta}),
\qquad
\eta_i = \boldsymbol{a}_i^\top \boldsymbol{x},
\end{equation}
where $\boldsymbol{a}_i^\top$ is typically a sparse row of the 
design matrix mapping the latent field to the linear predictor. 
A wide range of density functions can be used as a likelihood 
$p(\cdot)$. The only technical requirement is that it needs to have 
first and second order derivatives with respect to the latent field. 
The \pkg{INLA} engine supports a large catalogue of likelihood 
families as well as user-defined likelihoods; \textsf{PyINLA} 
currently exposes the most commonly used subset 
(see Section~\ref{sec:families}).

\textbf{Latent field prior.} Conditional on $\boldsymbol{\theta}$,
\begin{equation}
\boldsymbol{x}\mid \boldsymbol{\theta} \sim \mathcal{N}\!\left(
\boldsymbol{\mu}(\boldsymbol{\theta}),\ 
\boldsymbol{Q}(\boldsymbol{\theta})^{-1}\right),
\end{equation}
where $\boldsymbol{Q}(\boldsymbol{\theta})$ is a precision matrix
(the inverse of the covariance matrix). 
In most structured models, $\boldsymbol{Q}(\boldsymbol{\theta})$ is 
sparse, encoding conditional independence relationships that are 
central to computational efficiency \citep{rue2005gaussian}.

\textbf{Hyperparameter prior.} The hyperparameters are the 
extra likelihood parameters (not in the linear predictor) 
such as variance, dispersion, zero-inflation and 
the parameters on $\boldsymbol{Q}(\boldsymbol{\theta})$.
They are assigned a prior distribution $p(\boldsymbol{\theta})$, 
often including penalized complexity priors for variance and 
correlation parameters \citep{simpson2017penalising}. PC priors are specified through intuitive probability statements rather than abstract distributional parameters. For example, 
$P(\sigma > 1) = 0.01$ encodes the belief that a standard deviation exceeding 1 is unlikely, shrinking toward the simpler base model with no random effect.

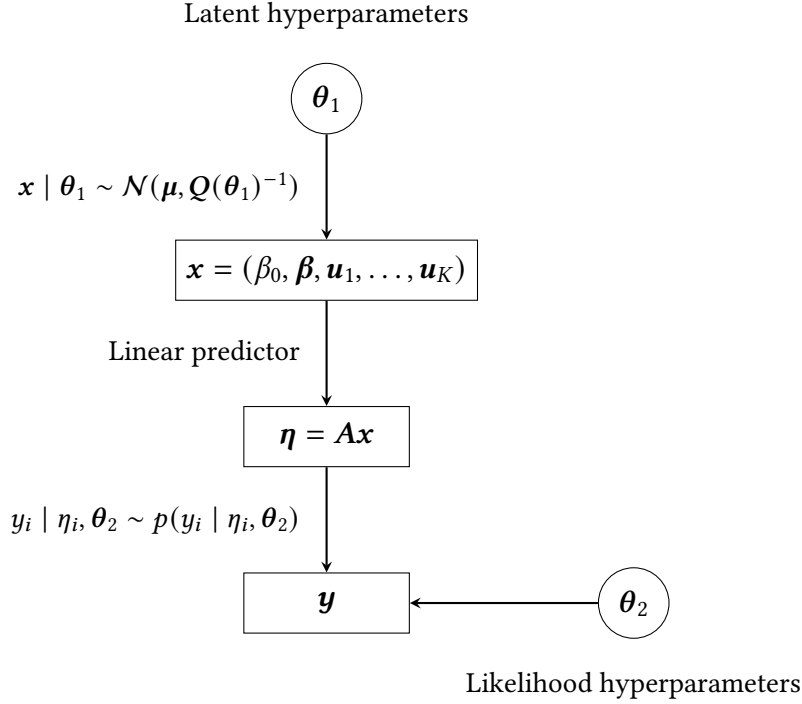
\begin{figure}[t]
\centering
\begin{tikzpicture}[
    node distance=1.4cm and 1.5cm,
    box/.style={rectangle, draw, minimum width=2.2cm, minimum height=0.8cm, align=center},
    param/.style={circle, draw, minimum size=0.9cm, align=center},
    arrow/.style={->, >=stealth, thick}
]
\node[param] (theta1) {$\boldsymbol{\theta}_1$};
\node[above=0.3cm of theta1, font=\small] {Latent hyperparameters};

\node[box, below=1.4cm of theta1] (x) {$\boldsymbol{x} = (\beta_0, \boldsymbol{\beta}, \boldsymbol{u}_1, \ldots, \boldsymbol{u}_K)$};

\node[box, below=1.4cm of x] (eta) {$\boldsymbol{\eta} = \boldsymbol{A}\boldsymbol{x}$};

\node[box, below=1.4cm of eta] (y) {$\boldsymbol{y}$};

\node[param, right=2.5cm of y] (theta2) {$\boldsymbol{\theta}_2$};
\node[below=0.3cm of theta2, font=\small] {Likelihood hyperparameters};

\draw[arrow] (theta1) -- (x) node[midway, left, font=\small, xshift=-2mm] {$\boldsymbol{x} \mid \boldsymbol{\theta}_1 \sim \mathcal{N}(\boldsymbol{\mu}, \boldsymbol{Q}(\boldsymbol{\theta}_1)^{-1})$};
\draw[arrow] (x) -- (eta) node[midway, left, font=\small, xshift=-2mm] {Linear predictor};
\draw[arrow] (eta) -- (y) node[midway, left, font=\small, xshift=-2mm] 
    {$y_i \mid \eta_i, \boldsymbol{\theta}_2 \sim 
    p(y_i \mid \eta_i, \boldsymbol{\theta}_2)$};
\draw[arrow] (theta2) -- (y);
\end{tikzpicture}
\caption{Hierarchical structure of a latent Gaussian model. Hyperparameters split into latent hyperparameters 
$\boldsymbol{\theta}_1$, which govern the precision structure of the latent field $\boldsymbol{x}$, and likelihood hyperparameters $\boldsymbol{\theta}_2$ (e.g., observation noise variance), with $\boldsymbol{\theta} = (\boldsymbol{\theta}_1, \boldsymbol{\theta}_2)$. The linear predictor $\boldsymbol{\eta}$ links $\boldsymbol{x}$ to 
observations $\boldsymbol{y}$ through the likelihood $p(\cdot)$.}
\label{fig:lgm-structure}
\end{figure}

Figure~\ref{fig:lgm-structure} illustrates the hierarchical 
structure of an LGM.

In \textsf{PyINLA}, each \code{family} comes with a default link 
function, otherwise it can be set explicitly. Fixed effects are 
specified via \code{fixed} and random effects via \code{random} 
(see Section~\ref{sec:pyinla}).

\subsection{Inference targets and outputs}

INLA targets posterior marginal distributions for each element of 
the latent field and hyperparameters:
\begin{equation}
p(x_j\mid \boldsymbol{y}) = \int p(x_j\mid \boldsymbol{\theta},
\boldsymbol{y})\,p(\boldsymbol{\theta}\mid \boldsymbol{y})\,
d\boldsymbol{\theta},
\qquad
p(\theta_k\mid \boldsymbol{y}) = \int p(\boldsymbol{\theta}\mid 
\boldsymbol{y})\,d\boldsymbol{\theta}_{-k}.
\end{equation}
These marginal distributions are non-Gaussian, and the Laplace 
method exploits the Gaussian prior on $\boldsymbol{x}$ to approximate them accurately and efficiently.
Once obtained, posterior summaries such as means, standard deviations, and credible intervals are computed directly from these marginals, and returned as \pkg{pandas} DataFrames (see Section~\ref{sec:pyinla}).
INLA focuses on marginals rather than representing the full joint posterior explicitly; this design is a key contributor to its computational efficiency. When the full joint posterior is needed, Monte Carlo samples can be drawn efficiently after the model fit \citep{joint2021}, as described in Section~\ref{sec:pyinla}.

\subsection{Nested Laplace approximation and numerical integration}

INLA proceeds by combining (i) Laplace approximations for 
high-dimensional Gaussian latent structure with (ii) numerical 
integration over the hyperparameter space \citep{Rue2009, Rue2017}. 
At a high level:
\begin{enumerate}
\item Construct a Gaussian approximation to 
      $p(\boldsymbol{x}\mid \boldsymbol{\theta},\boldsymbol{y})$ 
      to compute a Laplace approximation to 
      $p(\boldsymbol{\theta}\mid\boldsymbol{y})$ and use it to 
      search for the posterior mode of $\boldsymbol{\theta}$.
\item Approximate the conditional marginals 
      $p(x_j\mid \boldsymbol{\theta},\boldsymbol{y})$, for a set 
      of support points $\boldsymbol{\theta}$ around its mode.
\item Integrate over $\boldsymbol{\theta}$ using the set of support 
      points with weights proportional to 
      $p(\boldsymbol{\theta}\mid\boldsymbol{y})$ to obtain 
      $p(x_j\mid \boldsymbol{y})$ and 
      $p(\theta_k\mid \boldsymbol{y})$.
\end{enumerate}
\textit{Note:} if the likelihood is Gaussian, no approximation is 
required in steps 1 and 2.

\subsection{Computational considerations and limitations}

The first step of INLA is an optimization to
locate the mode of the hyperparameter, and
uses an adaptive gradient estimation technique \citep{fattah2022smart}.
The \textbf{inla} program exploits sparsity in the latent precision matrix via Cholesky factorization 
and selective inversion \citep{rue2005gaussian}. 
When the precision matrix is sparse, these operations are fast, 
but they become a bottleneck if the precision matrix is dense. Prior work explored solutions 
for specific cases, including dense matrices \citep{abdul2025inla+} and 
large-scale spatiotemporal models \citep{gaedke2024integrated}. 
The \textbf{sTiles} library \citep{fattah2025stiles, fattah2025gpu} is being developed to provide efficient factorization for sparse structured matrices, with GPU acceleration planned 
for future releases.
The \textbf{inla} source code is available at 
\url{https://github.com/hrue/r-inla}. 

\section{The interface description}\label{sec:pyinla}

This section describes the design, architecture, and usage of \textsf{PyINLA},
a \proglang{Python}-native interface to the compiled \textbf{inla} program that enables model specification, execution, and result handling entirely within standard \proglang{Python} workflows.

\subsection{Installation and availability}

\textsf{PyINLA} can be installed from PyPI:
\begin{CodeChunk}
\begin{CodeInput}
pip install pyinla
\end{CodeInput}
\end{CodeChunk}
Optional extras are available for workflows that require additional 
functionality, for example mesh construction utilities for SPDE-based 
spatial models:
\begin{CodeChunk}
\begin{CodeInput}
pip install pyinla[fmesher]
\end{CodeInput}
\end{CodeChunk}
The \code{[fmesher]} extra provides \proglang{Python} wrappers around 
the \pkg{fmesher} library \citep{lindgren2023fmesher} for mesh generation 
and SPDE preprocessing. We recommend installation in an isolated 
environment (e.g., \code{conda}).

\textsf{PyINLA} is a standard \proglang{Python} package with no \proglang{R} dependency. The heavy lifting is done by the \textbf{inla} program, a compiled \proglang{C} binary that implements the core numerical routines, sparse matrix factorizations, Laplace approximations, and numerical integration over hyperparameters. This is the same engine that has powered \pkg{R-INLA} for over 15 years and has been validated across thousands of published analyses. \textsf{PyINLA} manages this binary transparently: it is automatically downloaded on first use if not already present, so no manual configuration is required.
The package runs directly in cloud notebook environments
such as Google Colab, making it suitable for teaching, reproducible
demonstrations, and collaborative research without local installation.
Full documentation, tutorials, 
and interactive examples are available at \url{https://pyinla.org}. 
To support reproducible research, the repository includes replication 
materials for all results reported in this manuscript. 
This paper describes \textsf{PyINLA} version 0.2.0.

\subsection{Package architecture}

\textsf{PyINLA} is organized into five layers:
\begin{enumerate}
  \item \textbf{User-facing API:} A high-level interface centered on 
        the \code{pyinla()} function, which accepts model specification, 
        data, and control options and returns a structured result object.
  \item \textbf{Model specification layer:} Translation of 
        \proglang{Python}-level model definitions into the internal 
        representation required by the INLA engine, including validation 
        and defaults.
  \item \textbf{Data preparation layer:} Conversion of common 
        \proglang{Python} objects, \pkg{pandas} DataFrames, \pkg{NumPy} 
        arrays, and \pkg{SciPy} sparse matrices, into formats consumed 
        by the engine.
  \item \textbf{Execution layer:} Orchestration of INLA runs, including 
        management of working directories, configuration generation, 
        invocation of the compiled engine, and error reporting.
  \item \textbf{Result collection layer:} Parsing of engine outputs and 
        construction of \proglang{Python} result objects that expose 
        posterior marginals, summaries, and diagnostics.
\end{enumerate}

\subsection{Basic usage}

A minimal example fitting a Gaussian linear model illustrates the primary workflow. The code below generates synthetic data, specifies the model, fits it with \code{pyinla()}, and prints the posterior summary of fixed effects.

\begin{CodeChunk}
\begin{CodeInput}
from pyinla import pyinla
import pandas as pd
import numpy as np

rng = np.random.default_rng(42)
n = 1000
x = rng.standard_normal(n)
y = -2.0 + 1.5 * x + rng.standard_normal(n) * 0.5
data = pd.DataFrame({"y": y, "x": x})

model = {"response": "y", "fixed": ["1", "x"]}
result = pyinla(model=model, family="gaussian", data=data)

print(result.summary_fixed)
\end{CodeInput}

\begin{CodeOutput}
                 mean      sd  0.025quant  0.5quant  0.975quant     mode       kld
(Intercept)    -2.041   0.016      -2.072    -2.041      -2.009   -2.041  1.21e-11
x               1.496   0.016       1.464     1.496       1.528    1.496  1.05e-11
\end{CodeOutput}
\end{CodeChunk}

\noindent Each row reports the posterior mean, standard deviation,
quantiles, and mode for a fixed effect. The \code{kld} column is useful for non-Gaussian likelihood as it is the Kullback--Leibler divergence between the Gaussian and simplified Laplace approximation to each marginal; values near zero indicate that the Gaussian approximation is adequate. In this example it is zero as no Gaussian approximation is required when the family is Gaussian.

\begin{CodeChunk}
\begin{CodeInput}
import matplotlib.pyplot as plt

marg_x = result.marginals_fixed["x"]
plt.figure(figsize=(5, 3))
plt.plot(marg_x[:, 0], marg_x[:, 1])
plt.axvline(1.5, color="red", linestyle="--", label="True value")
plt.xlabel("x")
plt.ylabel("Density")
plt.legend()
plt.tight_layout()
\end{CodeInput}
\end{CodeChunk}

\begin{figure}[t]
\centering
\includegraphics[width=0.6\textwidth]{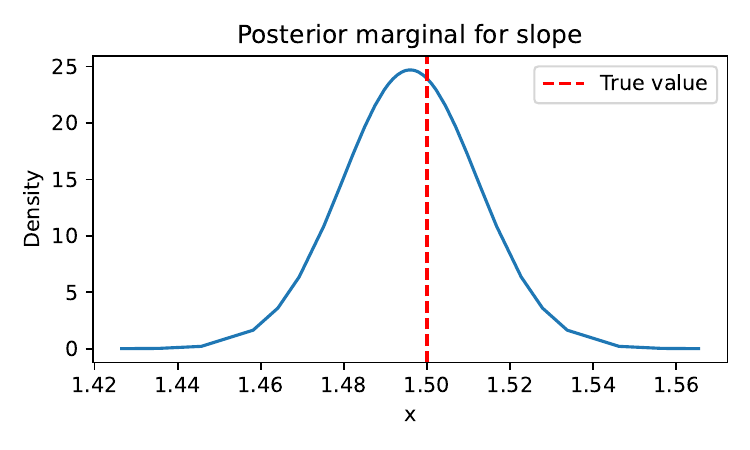}
\caption{Posterior marginal density for the slope parameter. The dashed red line indicates the true value (1.5) used in data generation.}
\label{fig:marginal-example}
\end{figure}

\subsection{Model specification}\label{sec:model-spec}

A statistical model in \textsf{PyINLA} is specified as a \proglang{Python} dictionary.
To motivate this representation, we build up from a simple example to the general case,
showing at each step how the dictionary maps to the underlying mathematical model.

\subsubsection{From equation to dictionary: a progressive construction}

\paragraph{Step 1: A linear model.}
Consider the simplest case: a Gaussian linear model with one covariate,
\begin{equation}\label{eq:step1}
  y_i = \beta_0 + \beta_1 x_i + \varepsilon_i,
  \qquad \varepsilon_i \sim \mathcal{N}(0, \sigma^2).
\end{equation}
In \textsf{PyINLA}, this model is specified as:
\begin{CodeChunk}
\begin{CodeInput}
model = {"response": "y", "fixed": ["1", "x"]}
result = pyinla(model=model, family="gaussian", data=data)
\end{CodeInput}
\end{CodeChunk}
The dictionary has two keys. The \code{"response"} key names the column in \code{data} that contains the observed values $y_i$. The \code{"fixed"} key lists the terms in the linear predictor: \code{"1"} denotes the intercept $\beta_0$, and \code{"x"} denotes the covariate column whose coefficient is $\beta_1$. The argument \code{family="gaussian"} specifies Gaussian observations with an identity link, meaning the linear predictor equals the conditional mean directly: $\eta_i = \mathrm{E}[y_i] = \beta_0 + \beta_1 x_i$.

\paragraph{Step 2: Changing the likelihood and link function.}
Now suppose the response is a count variable, and we wish to model it as Poisson:
\begin{equation}\label{eq:step2}
  y_i \sim \mathrm{Poisson}(\mu_i),
  \qquad \log(\mu_i) = \beta_0 + \beta_1 x_i.
\end{equation}
The only change in the code is the \code{family} argument:
\begin{CodeChunk}
\begin{CodeInput}
model = {"response": "y", "fixed": ["1", "x"]}
result = pyinla(model=model, family="poisson", data=data)
\end{CodeInput}
\end{CodeChunk}
The dictionary is identical, because the linear predictor $\eta_i = \beta_0 + \beta_1 x_i$ has not changed. What changed is how $\eta_i$ relates to the observations: the Poisson family uses a \emph{log link}, so $\eta_i = \log(\mu_i)$, or equivalently $\mu_i = \exp(\eta_i)$.

Each likelihood family carries a default \emph{link function} $g(\cdot)$ that maps the expected response $\mu_i = \mathrm{E}[y_i]$ to the linear predictor:
\begin{equation}\label{eq:link}
  \eta_i = g(\mu_i).
\end{equation}
The link function ensures that the linear predictor, which takes values on the entire real line, maps to the appropriate domain for each distribution (e.g., positive reals for counts, the unit interval for proportions). Table~\ref{tab:link-functions} lists the commonly used families and their default links.

\begin{table}[t]
\centering
\caption{Commonly used likelihood families in \textsf{PyINLA}. Each family defines a distribution for the observed data and a default link function $g(\cdot)$ relating the linear predictor $\eta_i$ to the expected response $\mu_i$.}
\label{tab:link-functions}
\small
\begin{tabular}{lllll}
\toprule
\code{family} & Distribution & Default link $g(\mu_i)$ & Inverse: $\mu_i =$ & Domain of $\mu_i$ \\
\midrule
\code{"gaussian"} & $\mathcal{N}(\mu_i, \sigma^2)$ & $\eta_i = \mu_i$ (identity) & $\eta_i$ & $(-\infty, \infty)$ \\
\code{"poisson"} & $\mathrm{Poisson}(\mu_i)$ & $\eta_i = \log \mu_i$ & $e^{\eta_i}$ & $(0, \infty)$ \\
\code{"binomial"} & $\mathrm{Bin}(n_i, p_i)$ & $\eta_i = \log\!\frac{p_i}{1-p_i}$ (logit) & $\frac{e^{\eta_i}}{1+e^{\eta_i}}$ & $(0, 1)$ \\
\code{"nbinomial"} & $\mathrm{NegBin}(\mu_i, \phi)$ & $\eta_i = \log \mu_i$ & $e^{\eta_i}$ & $(0, \infty)$ \\
\code{"gamma"} & $\mathrm{Gamma}(\mu_i, \phi)$ & $\eta_i = \log \mu_i$ & $e^{\eta_i}$ & $(0, \infty)$ \\
\code{"beta"} & $\mathrm{Beta}(\mu_i, \phi)$ & $\eta_i = \log\!\frac{\mu_i}{1-\mu_i}$ (logit) & $\frac{e^{\eta_i}}{1+e^{\eta_i}}$ & $(0, 1)$ \\
\bottomrule
\end{tabular}
\end{table}

The link function is handled internally by \textsf{PyINLA}: users specify the model on the linear predictor scale. Posterior summaries for fitted values are reported on the linear predictor scale; to obtain results on the response scale (e.g., expected counts or probabilities), the inverse link can be applied using \code{pyinla.tmarginal()}, as illustrated in Section~\ref{sec:marginal-utils}.

\paragraph{Step 3: Adding random effects.}
Suppose we extend the Poisson model with a group-level random intercept:
\begin{equation}\label{eq:step3}
  y_i \sim \mathrm{Poisson}(\mu_i),
  \qquad \log(\mu_i) = \beta_0 + \beta_1 x_i + u_{g(i)},
  \qquad u_j \stackrel{\mathrm{iid}}{\sim} \mathcal{N}(0, \sigma_u^2),
\end{equation}
where $g(i) \in \{1, \ldots, J\}$ is the group membership of observation $i$.
The \code{data} DataFrame must contain a column \code{group} with integer
indices identifying each observation's group. The dictionary gains a
\code{"random"} key:
\begin{CodeChunk}
\begin{CodeInput}
model = {
    "response": "y",
    "fixed": ["1", "x"],
    "random": [
        {"id": "group", "model": "iid"}
    ]
}
result = pyinla(model=model, family="poisson", data=data)
\end{CodeInput}
\end{CodeChunk}
The \code{"id": "group"} entry tells \textsf{PyINLA} to look up the column \code{group} in the data;
its values determine which observations share the same random effect level $u_j$.
The \code{"model": "iid"} entry specifies that the $u_j$ are exchangeable (independent, identically distributed) Gaussian, as in Equation~\eqref{eq:step3}.

\paragraph{Step 4: Multiple structured components.}
Models commonly include several random effects with different dependence structures. For example, a model with both group effects and a smooth temporal trend:
\begin{equation}\label{eq:step4}
  \log(\mu_i) = \beta_0 + \beta_1 x_i + u_{g(i)} + f_{t(i)},
\end{equation}
where $u_j \stackrel{\mathrm{iid}}{\sim} \mathcal{N}(0, \sigma_u^2)$ and $f_t$ follows a first-order random walk,
$f_t - f_{t-1} \sim \mathcal{N}(0, \sigma_f^2)$:
\begin{CodeChunk}
\begin{CodeInput}
model = {
    "response": "y",
    "fixed": ["1", "x"],
    "random": [
        {"id": "group", "model": "iid"},
        {"id": "time",  "model": "rw1", "constr": True}
    ]
}
result = pyinla(model=model, family="poisson", data=data)
\end{CodeInput}
\end{CodeChunk}
Each entry in the \code{"random"} list adds one latent component to the linear predictor. The \code{"constr": True} option imposes a sum-to-zero constraint ($\sum_t f_t = 0$) so that the random walk is identifiable alongside the intercept.

\paragraph{Summary.}
The general pattern follows Equation~\eqref{eq:general-pattern-preview}: 
the dictionary specifies the right-hand side (fixed + random) and the 
\code{family} argument specifies both the link function $g$ and the 
observation distribution $p$.

Unlike formula-based interfaces common in statistical software (e.g., \pkg{patsy} or \pkg{statsmodels} in \proglang{Python}, or the formula syntax in \proglang{R}), \textsf{PyINLA} uses plain dictionaries. This design supports programmatic model construction, for example when iterating over candidate model structures in a loop, reading specifications from configuration files, or building models conditionally based on data properties.

\subsubsection{Fixed effects}

Fixed effects are specified as a list of column names from the \code{data} DataFrame. The intercept is included by specifying \code{"1"}; omitting it from the list excludes the intercept:
\begin{CodeChunk}
\begin{CodeInput}
model = {"response": "y", "fixed": ["1", "x1", "x2", "x3"]}
\end{CodeInput}
\end{CodeChunk}
This corresponds to $\eta_i = \beta_0 + \beta_1 x_{1i} + \beta_2 x_{2i} + \beta_3 x_{3i}$.
The regression coefficients receive default Gaussian priors with low precision (wide, weakly informative priors).

\subsubsection{Likelihood families}\label{sec:families}

\textsf{PyINLA} supports families for continuous responses (\code{"gaussian"}, \code{"gamma"}, \code{"beta"}), discrete responses (\code{"poisson"}, \code{"binomial"}, \code{"nbinomial"}), and survival data (\code{"weibullsurv"}, \code{"exponentialsurv"}). The default link functions were listed in Table~\ref{tab:link-functions}.

Common additional arguments include:
\begin{itemize}
  \item \code{E}: expected counts for Poisson models with an offset, so that $\log(\mu_i) = \log(E_i) + \eta_i$ and the model estimates the relative risk $\mu_i / E_i$,
  \item \code{Ntrials}: the number of trials $n_i$ for binomial data, so that $y_i \sim \mathrm{Bin}(n_i, p_i)$,
  \item \code{weights} and \code{scale}: observation weights and scale vectors.
\end{itemize}

\begin{CodeChunk}
\begin{CodeInput}
# Poisson model for case counts with expected counts E_i
# Model: cases_i ~ Poisson(E_i * theta_i), log(theta_i) = beta_0 + beta_1 * x_i
result = pyinla(
    model={"response": "cases", "fixed": ["1", "x"]},
    family="poisson",
    data=data,
    E=data["expected"].to_numpy()
)

# Binomial model for successes out of n_i trials
# Model: successes_i ~ Bin(n_i, p_i), logit(p_i) = beta_0 + beta_1 * treatment_i
result = pyinla(
    model={"response": "successes", "fixed": ["1", "treatment"]},
    family="binomial",
    data=data,
    Ntrials=data["n_trials"].to_numpy()
)
\end{CodeInput}
\end{CodeChunk}

\subsubsection{Random effects (latent components)}

Random effects are specified under the \code{"random"} key as a list of component dictionaries. Each component requires:
\begin{itemize}
  \item \code{"id"}: a column name in the data whose values index the random effect levels,
  \item \code{"model"}: the type of prior dependence structure imposed on the effect.
\end{itemize}

Table~\ref{tab:latent-models} summarizes the available latent model types with their mathematical definitions.

\begin{table}[t]
\centering
\caption{Built-in latent model types in \textsf{PyINLA}. Each model defines a prior distribution over the random effect vector $\boldsymbol{u} = (u_1, \ldots, u_m)$.}
\label{tab:latent-models}
\small
\begin{tabular}{lll}
\toprule
\code{"model"} & Name & Prior structure \\
\midrule
\code{"iid"} & Exchangeable & $u_j \stackrel{\mathrm{iid}}{\sim} \mathcal{N}(0, \tau^{-1})$ \\
\code{"rw1"} & Random walk (order 1) & $u_t - u_{t-1} \sim \mathcal{N}(0, \tau^{-1})$ \\
\code{"rw2"} & Random walk (order 2) & $u_{t-1} - 2u_t + u_{t+1} \sim \mathcal{N}(0, \tau^{-1})$ \\
\code{"ar1"} & Autoregressive (order 1) & $u_t = \rho \, u_{t-1} + \varepsilon_t,\ \varepsilon_t \sim \mathcal{N}(0, \sigma^2_\varepsilon)$ \\
\code{"bym"} & BYM (areal spatial) & $b_i = u_i + v_i$;\ \ $u_i$: ICAR$(\tau_u)$,\ $v_i \stackrel{\mathrm{iid}}{\sim} \mathcal{N}(0, \tau_v^{-1})$ \\
\code{"bym2"} & BYM2 (areal spatial) & $b_i = \sigma(\sqrt{\phi}\, u^*_i + \sqrt{1-\phi}\, v^*_i)$ \\
\code{"spde"} & Mat\'ern field (SPDE) & Continuous spatial field via SPDE \\
\code{"generic0"} & Custom precision & User-supplied $\boldsymbol{Q}$: $\boldsymbol{u} \sim \mathcal{N}(\boldsymbol{0}, \tau^{-1}\boldsymbol{Q}^{-1})$ \\
\bottomrule
\end{tabular}
\end{table}

The \code{"iid"} model is the simplest: it assigns independent Gaussian priors to each level, suitable for group-level intercepts (e.g., subjects, regions, or items in a mixed model). The random walk models (\code{"rw1"}, \code{"rw2"}) produce smooth functions over ordered indices (time steps, ages, ordinal categories) by penalizing differences between consecutive values. The \code{"ar1"} model is appropriate for stationary time series where the correlation between $u_t$ and $u_{t+k}$ decays as $\rho^{|k|}$. The \code{"bym"} and \code{"spde"} models handle spatial dependence on lattices and continuous domains, respectively, and are illustrated in Sections~\ref{sec:ex-scottish} and~\ref{sec:ex-spde}.

In addition to \code{"id"} and \code{"model"}, each component dictionary accepts optional flags:
\begin{itemize}
  \item \code{"constr": True} imposes a sum-to-zero constraint ($\sum_j u_j = 0$), ensuring identifiability when the component is collinear with the intercept.
  \item \code{"scale.model": True} scales the precision matrix so that its generalized variance is one, making the precision hyperparameter comparable across models of different dimension or graph structure.
  \item \code{"cyclic": True} wraps the random walk so that the last level is adjacent to the first, suitable for periodic effects such as day-of-week or day-of-year seasonality.
\end{itemize}

\subsubsection{Prior specification}

Each latent component has hyperparameters (e.g., the precision $\tau$ for an IID effect, or the correlation $\rho$ for an AR1 process) that require prior distributions. These are specified using the \code{"hyper"} key within the random effect dictionary.

\textsf{PyINLA} supports penalized complexity (PC) priors \citep{simpson2017penalising}, which are specified through interpretable probability statements. For instance, the statement
\begin{equation}\label{eq:pc-prior}
  \Pr(\sigma > U) = \alpha
\end{equation}
sets a prior on the standard deviation $\sigma = 1/\sqrt{\tau}$ such that the probability of $\sigma$ exceeding a threshold $U$ is $\alpha$. In code:
\begin{CodeChunk}
\begin{CodeInput}
model = {
    "response": "y",
    "fixed": ["1", "x"],
    "random": [
        {
            "id": "group",
            "model": "iid",
            "hyper": {
                "prec": {
                    "prior": "pc.prec",
                    "param": [1.0, 0.01]  # P(sigma > 1.0) = 0.01
                }
            }
        }
    ]
}
\end{CodeInput}
\end{CodeChunk}
Here \code{"param": [1.0, 0.01]} encodes Equation~\eqref{eq:pc-prior} with $U = 1.0$ and $\alpha = 0.01$: the prior places only 1\% probability on the random effect standard deviation exceeding 1.0, which shrinks toward zero (the base model with no random effect). Table~\ref{tab:priors} lists the most commonly used prior families.

\begin{table}
\centering
\caption{Commonly used prior families in \textsf{PyINLA}. The \code{"prior"} string
is passed inside the \code{"hyper"} dictionary; \code{"param"} supplies the
corresponding parameters.}
\label{tab:priors}
\small
\begin{tabular}{lll}
\toprule
\code{"prior"} & Parameters & Description \\
\midrule
\code{"pc.prec"} & $[U, \alpha]$ & PC prior on $\sigma$: $P(\sigma > U) = \alpha$ \\
\code{"pc.cor1"} & $[U, \alpha]$ & PC prior on $\rho$ (base model $\rho = 1$) \\
\code{"pc.cor0"} & $[U, \alpha]$ & PC prior on $\rho$ (base model $\rho = 0$) \\
\code{"pc"} & $[U, \alpha]$ & Generic PC prior, e.g.\ BYM2 mixing $\phi$ \\
\code{"loggamma"} & $[a, b]$ & Log-Gamma on $\log\tau$ (shape $a$, rate $b$) \\
\code{"gaussian"} & $[\mu, \tau]$ & Gaussian on $\theta$ (mean $\mu$, precision $\tau$) \\
\code{"flat"} & --- & Improper constant (non-informative) \\
\code{"pc.dof"} & $[U, \alpha]$ & PC prior on degrees of freedom \\
\code{"betacorrelation"} & $[a, b]$ & Beta prior on a correlation \\
\code{"table:"} & grid & Tabulated log-density on $\theta$ \\
\bottomrule
\end{tabular}
\end{table}

\subsection{Result object}

The \code{pyinla()} function returns a \code{PyINLAresult} object that exposes posterior summaries, marginal densities, and model assessment quantities in \proglang{Python}-native formats.

\subsubsection{Posterior summaries}

Summaries for fixed effects, latent components, hyperparameters, and fitted values are returned as \pkg{pandas} DataFrames:
\begin{CodeChunk}
\begin{CodeInput}
result.summary_fixed
result.summary_random
result.summary_hyperpar
result.summary_fitted_values
\end{CodeInput}
\end{CodeChunk}

\subsubsection{Marginal posteriors and utility functions} \label{sec:marginal-utils}

Marginal posterior densities are represented as arrays of $(x, \mathrm{density})$ pairs. Utility functions are provided for evaluation, transformation, and summarization (see Appendix~\ref{app:marginals}):

\begin{CodeChunk}
\begin{CodeInput}
import pyinla
import numpy as np

marg = result.marginals_fixed["x"]

prob_positive = 1.0 - pyinla.pmarginal(0.0, marg)   # CDF (p)
hpd95 = pyinla.hpdmarginal(0.95, marg)              # HPD interval

marg_or = pyinla.tmarginal(np.exp, marg)  # transform (t) marginal
summary = pyinla.zmarginal(marg)          # summary statistics (z)
\end{CodeInput}
\end{CodeChunk}

\subsection{Control options}\label{sec:control}

The \code{control} argument is an optional dictionary that adjusts computation and prior settings beyond the defaults. It is organized into named groups, each controlling a different aspect of the fit. The most commonly used groups are:

\begin{itemize}
  \item \code{"compute"}: toggles for additional output.
    \code{"dic"}, \code{"waic"}, and \code{"cpo"} enable the corresponding model-comparison criteria;
    \code{"config"} stores internal configurations needed by \code{posterior\_sample()};
    \code{"return\_marginals"} requests full marginal densities for fitted values.
  \item \code{"fixed"}: default prior precision for fixed effects.
    \code{"prec"} sets the precision (inverse variance) of the zero-mean Gaussian prior on regression coefficients, and
    \code{"prec.intercept"} sets it for the intercept specifically.
    Lower values give wider, less informative priors.
\end{itemize}

\noindent All keys are optional; unspecified settings retain their defaults.

\subsection{Model diagnostics}

When requested via \code{control}, \textsf{PyINLA} computes diagnostics commonly used with INLA workflows, including DIC, WAIC, CPO, and an approximation to the marginal likelihood:
\begin{CodeChunk}
\begin{CodeInput}
result = pyinla(
    model=model,
    family="gaussian",
    data=data,
    control={"compute": {"dic": True, "waic": True, "cpo": True}}
)

dic = result.dic
waic = result.waic
mlik = result.mlik
\end{CodeInput}
\end{CodeChunk}

\subsection{Posterior sampling}

Although INLA focuses on marginal posteriors, posterior sampling is useful for derived quantities depending jointly on multiple components. \textsf{PyINLA} provides three sampling functions:
\begin{itemize}
  \item \code{posterior\_sample(n, result, seed)} draws \code{n} joint samples from the approximate posterior of the latent field and hyperparameters. The \code{config} flag must be enabled in \code{control} (see Section~\ref{sec:control}).
  \item \code{posterior\_sample\_eval(fun, samples)} extracts or transforms the samples. When \code{fun} is a string (e.g., \code{"x"}), it returns the samples for the named fixed effect; when \code{fun} is a callable, it is applied to each sample to compute a derived quantity.
  \item \code{hyperpar\_sample(n, result)} draws \code{n} samples from the approximate marginal posterior of the hyperparameters only.
\end{itemize}

\begin{CodeChunk}
\begin{CodeInput}
result = pyinla(
    model=model,
    family="gaussian",
    data=data,
    control={"compute": {"config": True}}
)

samples = pyinla.posterior_sample(n=1000, result=result, seed=42)
x_samples = pyinla.posterior_sample_eval("x", samples)
exp_x = pyinla.posterior_sample_eval(lambda x, **_: np.exp(x), samples)
hyper_samples = pyinla.hyperpar_sample(n=1000, result=result)
\end{CodeInput}
\end{CodeChunk}

\subsection{Robust execution}

\textsf{PyINLA} includes automatic fallback strategies for numerical difficulties. If default settings encounter problems (e.g., instability or non-convergence in optimization), the package retries with more conservative settings such as alternative approximation modes, modified integration strategies, or stabilization options for ill-conditioned systems. This behavior is controlled by the \code{safe} argument (default: \code{True}).

\subsection{Supported features and ongoing development}

\textsf{PyINLA} supports commonly used likelihood families, latent components (exchangeable, temporal, spatial, and generic), posterior summaries and marginals, key diagnostics, and posterior sampling. Coverage continues to expand as additional components are tested and stabilized. The package uses the same compiled \textbf{inla} engine described in Section~\ref{sec:inla}, ensuring numerical accuracy of the underlying computations.

Full documentation of supported features is available at the package website.

\section{Examples}\label{sec:examples}

This section illustrates \textsf{PyINLA} through four applications of increasing structural complexity. The first applies a hierarchical Poisson model with exchangeable random effects to football match prediction, demonstrating the core generalized linear mixed model workflow. The second uses the scaled BYM model \citep{riebler2016intuitive} for areal disease mapping on the Scottish lip cancer dataset, introducing spatially structured random effects on a lattice. The third uses the SPDE approach \citep{lindgren2011explicit} for geostatistical temperature modeling with continuous spatial coordinates. The fourth demonstrates time series forecasting with uncertainty quantification, comparing \textsf{PyINLA} against the neural network-based \pkg{NeuralProphet} tool \citep{triebe2021neuralprophet}. Together, these examples progress from standard GLMMs through lattice and geostatistical spatial models to structured temporal models. Additional examples are available at \url{https://pyinla.org}. 

\subsection{Sports analytics: football match prediction}
\label{sec:ex-football}

This example demonstrates a Poisson generalized linear mixed model (GLMM) 
with crossed random effects for predicting football match outcomes. We 
compare PyINLA to PyMC \citep{Salvatier2016} with the No-U-Turn 
Sampler (NUTS) to validate posterior inference accuracy and quantify 
computational speedup.

\subsubsection{Data}

The dataset comprises 313 played matches (of 380 scheduled) from the 2018--2019 English Premier League
season involving 20 teams. Each match is restructured into two observations
(one per team), yielding $2 \times 313 = 626$ rows in long format:

\begin{CodeInput}
import pandas as pd
from pyinla import pyinla

# Load and restructure data
rows = []
for _, match in played_df.iterrows():
    rows.append({'goals': match['home_goals'], 
                 'attack': team_to_id[match['home_team']],
                 'defense': team_to_id[match['away_team']], 'home': 1})
    rows.append({'goals': match['away_goals'],
                 'attack': team_to_id[match['away_team']],
                 'defense': team_to_id[match['home_team']], 'home': 0})
data = pd.DataFrame(rows)
\end{CodeInput}

\subsubsection{Implementation}

The model specification uses IID random effects for attack and defense 
abilities:

\begin{CodeInput}
model = {
    'response': 'goals',
    'fixed': ['1', 'home'],
    'random': [
        {'id': 'attack', 'model': 'iid',
         'hyper': {'prec': {'prior': 'loggamma', 'param': [1, 0.01]}}},
        {'id': 'defense', 'model': 'iid',
         'hyper': {'prec': {'prior': 'loggamma', 'param': [1, 0.01]}}}
    ]
}

result = pyinla(model=model, family='poisson', data=data,
                control={'compute': {'dic': True, 'mlik': True}})
\end{CodeInput}

\begin{CodeOutput}
pyINLA fitting time: 0.24 seconds

Fixed effects:
                 mean        sd  0.025quant  0.975quant
(Intercept)  0.150466  0.100775   -0.050613    0.346319
home         0.237082  0.067770    0.104192    0.369971

Hyperparameters:
                            mean         sd  0.025quant  0.975quant
Precision for attack   12.408061   5.102970    5.216360   24.937430
Precision for defense  22.751203  11.570014    8.076030   52.307311
\end{CodeOutput}

The home-advantage coefficient $\hat{\beta} = 0.237$ corresponds to a
multiplicative increase in expected goals of $e^{0.237} \approx 1.27$,
indicating the home team scores approximately 27\% more goals than at
a neutral venue. The 95\% credible interval $[0.104, 0.370]$ lies entirely
above zero, providing strong evidence for a home-field effect.

\subsubsection{Comparison with MCMC}

To validate PyINLA's accuracy, we fit the identical model using PyMC with 
NUTS (4 chains, 25,000 draws each after 5,000 tuning iterations):

\begin{table}[t]
\centering
\caption{Timing and parameter comparison between PyINLA and MCMC (PyMC/NUTS)
for the football prediction model. PyINLA achieves a ${\sim}90\times$ speedup while
producing nearly identical posterior estimates.}
\label{tab:football-comparison}
\begin{tabular}{lrrrrr}
\hline
& \multicolumn{2}{c}{PyINLA} & \multicolumn{2}{c}{MCMC} & \\
Parameter & Mean & SD & Mean & SD & $|\Delta|$ \\
\hline
\multicolumn{6}{l}{\textit{Timing}} \\
\quad CPU time (s) & \multicolumn{2}{c}{0.24} & \multicolumn{2}{c}{21.74} & 92$\times$ \\
\hline
\multicolumn{6}{l}{\textit{Fixed effects}} \\
\quad Intercept & 0.150 & 0.101 & 0.150 & 0.100 & 0.001 \\
\quad Home & 0.237 & 0.068 & 0.237 & 0.068 & 0.000 \\
\hline
\multicolumn{6}{l}{\textit{Hyperparameters}} \\
\quad $\tau_{\text{attack}}$ & 12.41 & 5.10 & 12.41 & 5.06 & 0.00 \\
\quad $\tau_{\text{defense}}$ & 22.75 & 11.57 & 22.54 & 11.07 & 0.21 \\
\hline
\end{tabular}
\end{table}

Table~\ref{tab:football-comparison} reports wall-clock times for model
fitting.\footnote{All timings in this paper were obtained on a single
core of an Intel Core Ultra 7 155H (4.8\,GHz) with 64\,GB RAM, running
Ubuntu 24.04 and Python 3.12. MCMC timings reflect total sampling time
across all chains run sequentially; parallel execution would reduce
wall-clock time but not total CPU time.}
PyINLA completes inference in 0.24 seconds compared to 21.74 seconds
for MCMC (4 chains $\times$ 25{,}000 draws, run sequentially),
a speedup of approximately $92\times$ in CPU time. Fixed effect estimates agree to three decimal places, while
hyperparameter estimates differ by at most 0.21, well within posterior
uncertainty. The Pearson correlation across all 40 random effects (20 attack 
+ 20 defense) is $r = 1.0000$.

\begin{figure}[t!]
\centering
\includegraphics[width=\textwidth]{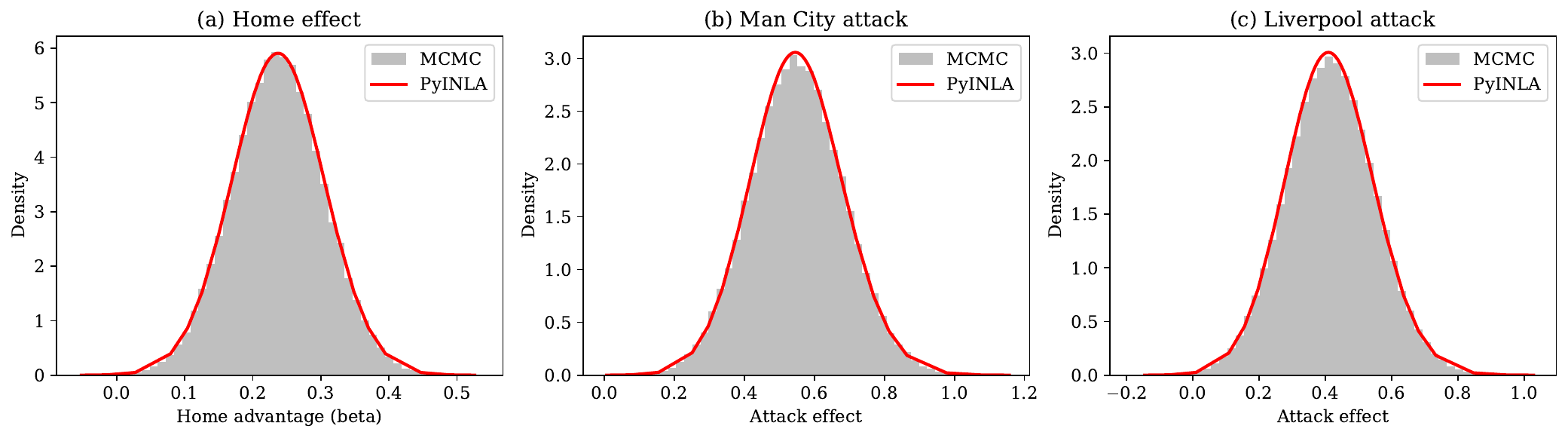}
\caption{Validation of PyINLA against MCMC. Left: INLA marginal density (red curve) overlaid on MCMC histogram (grey) for the home-advantage parameter. Center/Right: Attack effects for Manchester City and Liverpool. 
The close agreement confirms that the INLA approximation is accurate for this model class.}
\label{fig:football-validation}
\end{figure}

Figure~\ref{fig:football-validation} overlays PyINLA marginal densities on 
MCMC histograms for the home-advantage parameter and selected team effects, 
demonstrating excellent agreement between the two methods.

\subsubsection{Predictive simulation}

\begin{figure}[t!]
\centering
\includegraphics[width=\textwidth]{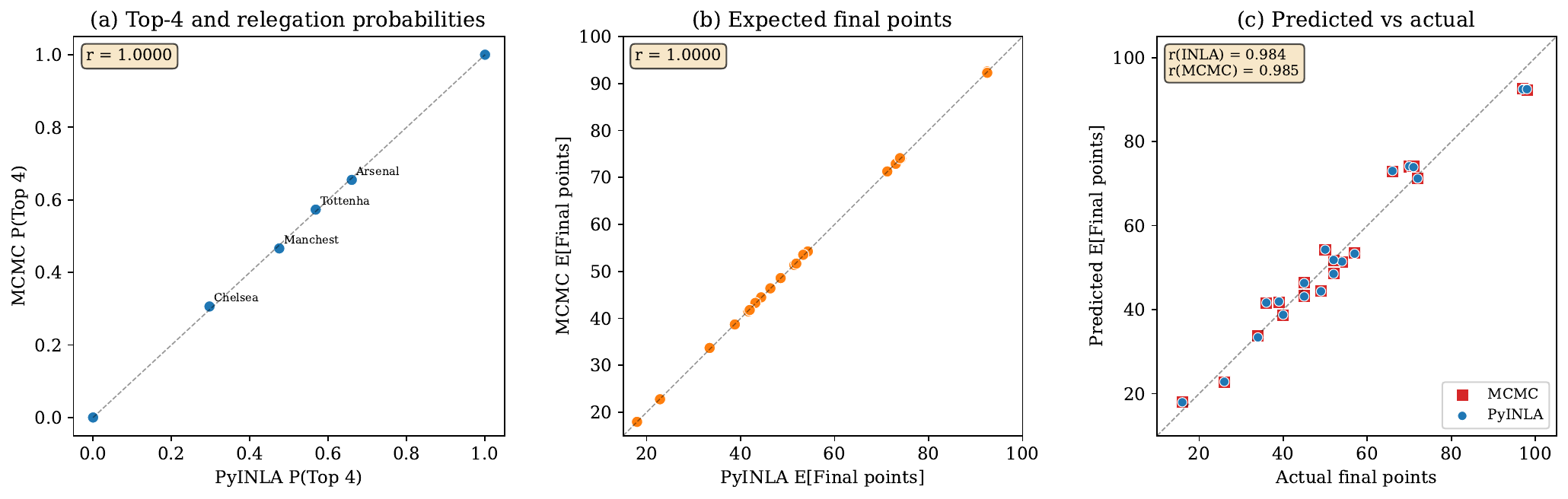}
\caption{Season simulation results. (a) Top-4 qualification probability: 
PyINLA vs MCMC ($r = 0.9999$). (b) Expected final points ($r = 0.9998$). 
(c) Predicted vs actual final points for both methods. Both approaches 
correctly identify Liverpool and Manchester City as title contenders and 
Huddersfield, Fulham, and Cardiff as relegation candidates.}
\label{fig:football-predictions}
\end{figure}

Using \code{posterior\_sample()} to draw 1,000 joint posterior samples, we simulate complete season outcomes for the 67 unplayed matches. 
Figure~\ref{fig:football-predictions} compares PyINLA and MCMC predictive distributions for top-4 qualification probabilities and expected final points.

Both methods correctly identify Liverpool and Manchester City as title 
contenders (P(Top 4) $\approx 1.0$) and Huddersfield, Fulham, and Cardiff 
as relegation candidates. The correlation between PyINLA and MCMC 
predictions exceeds 0.999 for all derived quantities, confirming that the 
INLA approximation preserves the information needed for downstream 
decision-making.

In summary, posterior means agree between PyINLA and MCMC to within $10^{-3}$ for all fixed and random effects, with a Pearson correlation of $r = 1.0$. PyINLA completes inference in under one second, achieving a 92$\times$ speedup over MCMC with 100,000 posterior draws. Simulated season outcomes, including qualification probabilities and expected points, are virtually identical between methods.

\paragraph{Replication materials.} The complete replication script is available as supplementary material (\texttt{football.py}).

\subsection{Time series forecasting: web traffic prediction} \label{sec:ex-timeseries}

This example demonstrates time series forecasting with uncertainty quantification, 
comparing PyINLA to NeuralProphet \citep{triebe2021neuralprophet}, a widely-used 
neural network-based forecasting tool. We use the Peyton Manning Wikipedia 
page views dataset (2007--2016), a standard benchmark in the forecasting 
literature containing 2964 daily observations of log-transformed page views.

\subsubsection{Model specification}

Let $y_t$ denote log page views at time $t$. We decompose the series as:
\begin{equation}\label{eq:ts-model}
y_t = \mu + f_{\text{trend}}(t) + f_{\text{week}}(w_t) + f_{\text{year}}(d_t) + u_t + \varepsilon_t,
\end{equation}
where $\mu$ is the intercept, $f_{\text{trend}}$ is a smooth trend modeled 
as a second-order random walk (RW2), $f_{\text{week}}$ captures day-of-week 
effects using a cyclic RW2 with period 7, $f_{\text{year}}$ captures yearly 
seasonality using a cyclic RW2 with period 365, $u_t$ follows an AR(1) process 
to model residual autocorrelation, and $\varepsilon_t \sim \mathcal{N}(0, \sigma^2_\varepsilon)$ 
is observation noise.

Calendar-aligned seasonality indices are computed directly from dates to ensure 
proper alignment:

\begin{CodeInput}
week_idx = dates.dt.dayofweek.values + 1   # Monday=1, ..., Sunday=7
year_idx = np.clip(dates.dt.dayofyear.values, 1, 365)  # Handle leap years
\end{CodeInput}

\subsubsection{Implementation}

We split the data into training (first 2599 days) and test (remaining 365 days) 
periods:

\begin{CodeInput}
import numpy as np
import pandas as pd
from pyinla import pyinla

# Load data
df = pd.read_csv("peyton_manning.csv", parse_dates=['ds'])
y = df['y'].values
dates = df['ds']

# Train/test split
N_TRAIN = 2599
y_train = y[:N_TRAIN]
n_total = len(y)

# Prepare data with NA for test period (predictions)
y_full = np.concatenate([y_train, np.full(n_total - N_TRAIN, np.nan)])

data = pd.DataFrame({
    'y': y_full,
    't': np.arange(1, n_total + 1),
    'week': (dates.dt.dayofweek.values + 1).astype(int),
    'year_day': np.clip(dates.dt.dayofyear.values, 1, 365).astype(int),
    't_ar': np.arange(1, n_total + 1)
})
\end{CodeInput}

The model specification includes PC priors encoding beliefs about component 
smoothness and autocorrelation strength:

\begin{CodeInput}
model = {
    'response': 'y',
    'fixed': ['1'],
    'random': [
        {'id': 't', 'model': 'rw2', 'scale.model': True,
         'hyper': {'prec': {'prior': 'pc.prec', 'param': [0.5, 0.01]}}},
        {'id': 'week', 'model': 'rw2', 'cyclic': True,
         'hyper': {'prec': {'prior': 'pc.prec', 'param': [1, 0.01]}}},
        {'id': 'year_day', 'model': 'rw2', 'cyclic': True,
         'hyper': {'prec': {'prior': 'pc.prec', 'param': [1, 0.01]}}},
        {'id': 't_ar', 'model': 'ar1',
         'hyper': {'prec': {'prior': 'pc.prec', 'param': [1, 0.01]},
                   'rho': {'prior': 'pc.cor1', 'param': [0.9, 0.9]}}}
    ]
}

result = pyinla(model=model, family='gaussian', data=data)
\end{CodeInput}

\subsubsection{Results}

Table~\ref{tab:ts-comparison} reports forecast accuracy on the 365-day test
period. NeuralProphet was run with both default and tuned hyperparameters,
with the best configuration (default settings) reported.
This comparison is illustrative rather than a formal benchmark:
NeuralProphet is a point-prediction tool (uncertainty intervals require
additional configuration), so the primary purpose is to demonstrate that
\textsf{PyINLA}'s structured additive model produces competitive point
forecasts while additionally providing calibrated uncertainty
quantification as a natural byproduct of Bayesian inference.

\begin{table}[t]
\centering
\caption{Forecast accuracy comparison on the Peyton Manning benchmark.}
\label{tab:ts-comparison}
\begin{tabular}{lrrrr}
\hline
Method & RMSE & MAE & 95\% CI Coverage & Avg.\ Width \\
\hline
NeuralProphet & 0.575 & 0.466 & --- & --- \\
PyINLA (no AR1) & 1.107 & 0.977 & 94.0\% & --- \\
PyINLA (with AR1) & \textbf{0.486} & \textbf{0.308} & 98.4\% & 2.75 \\
\hline
\end{tabular}
\end{table}

PyINLA with AR(1) achieves 15.5\% lower RMSE than NeuralProphet. The AR(1) 
component is critical: without it, PyINLA performs substantially worse 
(RMSE = 1.107), demonstrating the importance of modeling residual 
autocorrelation in this dataset.

The 95\% credible prediction interval coverage of 98.4\% indicates slightly conservative 
but well-calibrated uncertainty quantification. The estimated autocorrelation 
$\hat{\rho} \approx 0.65$ indicates moderate day-to-day persistence after 
accounting for trend and seasonality.

\begin{figure}[hb!]
\centering
\includegraphics[width=\textwidth]{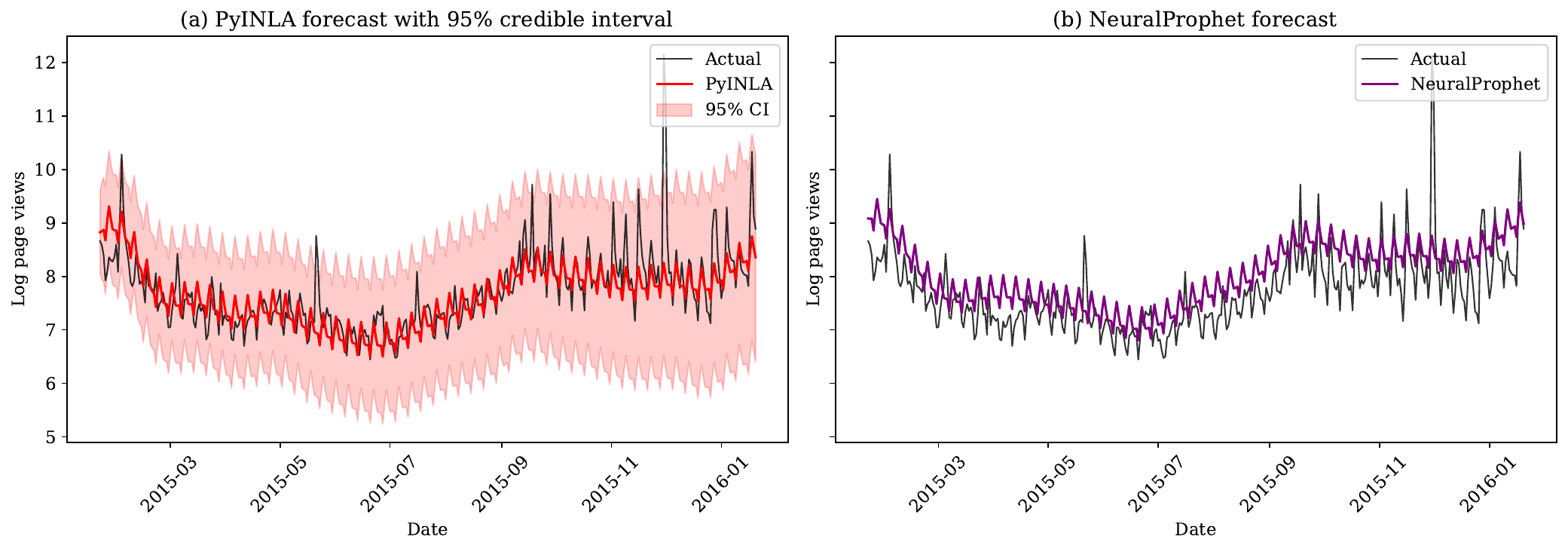}
\caption{Forecast comparison for Peyton Manning Wikipedia page views. 
Left: PyINLA forecast (red) with 95\% credible interval (shaded). 
Right: NeuralProphet forecast (purple). Black line shows actual test data.}
\label{fig:ts-forecast}
\end{figure}

Figure~\ref{fig:ts-forecast} shows the forecast comparison. PyINLA provides 
calibrated 95\% credible intervals that appropriately widen in the forecast 
period, while NeuralProphet provides only point predictions by default.

\begin{figure}[hb!]
\centering
\includegraphics[width=\textwidth]{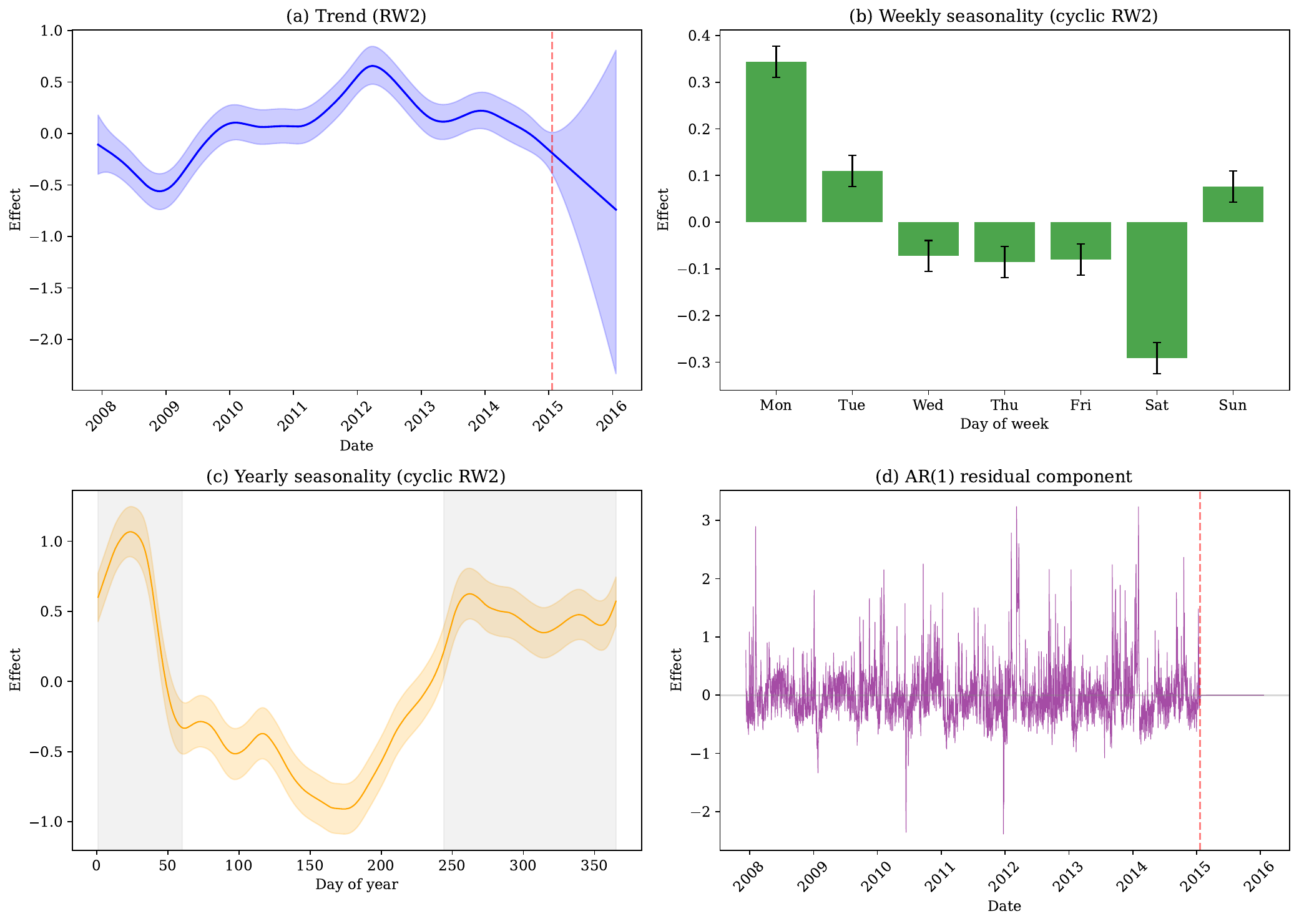}
\caption{Posterior decomposition of time series components. Top-left: trend 
(RW2). Top-right: weekly seasonality (cyclic RW2, period 7). Bottom-left: 
yearly seasonality (cyclic RW2, period 365). Bottom-right: AR(1) residuals. 
Shaded regions indicate 95\% credible intervals.}
\label{fig:ts-components}
\end{figure}

A key advantage of PyINLA is interpretable component decomposition with 
uncertainty. Figure~\ref{fig:ts-components} shows the posterior mean and 
95\% credible intervals for each component: the smooth trend captures 
long-term popularity changes, the weekly effect reveals weekday/weekend 
patterns, and the yearly effect shows NFL season dynamics (peaks during 
September--February).

\paragraph{Limitations.}
This comparison uses a single dataset with one train/test split. Results 
are illustrative rather than definitive; performance may differ on other 
datasets or with hyperparameter tuning. A comprehensive comparison would 
require multiple datasets and cross-validation.

\paragraph{Replication materials.}
The complete replication script is available as supplementary material 
(\texttt{timeseries.py}).

\subsection{Disease mapping: Scottish lip cancer}
\label{sec:ex-scottish}

This example demonstrates areal disease mapping using the classic BYM model
\citep{besag1991bayesian} with scaled precision matrix
(\texttt{scale.model=True}), applied to the Scottish lip cancer
dataset \citep{clayton1987empirical}. Scaling ensures that the precision
hyperparameters for the ICAR and IID components are comparable
\citep{sorbye2014scaling}. We compare PyINLA to PyMC with NUTS
to validate posterior inference on a spatially structured random effects model.

Note that the BYM2 reparameterisation \citep{riebler2016intuitive},
which uses a single total standard deviation and a mixing parameter, is also
available in \textsf{PyINLA} via \texttt{"model":~"bym2"} and is
recommended for new analyses; here we use the classic form to illustrate
separate hyperparameter estimation.

\subsubsection{Model specification}

Let $Y_i$ denote observed lip cancer cases in district $i$ with expected count
$E_i$ based on age-sex standardization. We model the relative risk $\theta_i$ as:
\begin{equation}\label{eq:bym-model}
Y_i \sim \text{Poisson}(E_i \cdot \theta_i), \qquad
\log \theta_i = \beta_0 + \beta_1 \cdot x_i + b_i,
\end{equation}
where $\beta_0$ is the intercept, $x_i$ is the proportion of the population
employed in agriculture, fishing, and forestry (AFF), and $b_i$ is a
spatially structured random effect following the scaled BYM decomposition:
\begin{equation}
b_i = \frac{\sigma_{\text{spatial}}}{\sqrt{s}} \cdot u_i + \sigma_{\text{iid}} \cdot v_i,
\end{equation}
where $u_i$ is an intrinsic conditional autoregressive (ICAR) component scaled
by the factor $s$ (the geometric mean of the marginal variances of the ICAR
precision matrix), $v_i \sim \mathcal{N}(0, 1)$ is an unstructured IID component,
and $\sigma_{\text{spatial}}$ and $\sigma_{\text{iid}}$ are the two standard
deviation hyperparameters. The scaling ensures that the two precision
hyperparameters are comparable, and the spatial fraction is derived post hoc
from posterior samples:
\begin{equation}
\phi = \frac{\sigma_{\text{spatial}}^2}{\sigma_{\text{spatial}}^2 + \sigma_{\text{iid}}^2}.
\end{equation}

\subsubsection{Data}

The dataset comprises 56 Scottish districts with lip cancer cases recorded
from 1975--1980. The AFF covariate captures occupational sun exposure, a known
risk factor for lip cancer:

\begin{CodeInput}
import pandas as pd
import numpy as np
from pyinla import pyinla

data = pd.read_csv("scotland_data.csv")
data['idarea'] = np.arange(1, len(data) + 1)
data['SIR'] = data['Y'] / data['E']

print(f"Districts: {len(data)}")
print(f"Total cases: {data['Y'].sum()}")
print(f"AFF range: {data['AFF'].min():.2f} to {data['AFF'].max():.2f}")
\end{CodeInput}

\begin{CodeOutput}
Districts: 56
Total cases: 536
AFF range: 0.00 to 0.24
\end{CodeOutput}

The adjacency graph defines spatial neighbors for each district.
Three singleton islands (Orkney, Shetland, Western Isles) have no neighbors
in the original graph and are connected to their nearest mainland district
to form a single connected component, yielding 120 edges across 56 nodes.

\subsubsection{Implementation}

The scaled BYM model is specified using penalized complexity (PC) priors
\citep{simpson2017penalising} for both precision hyperparameters. Each PC prior
encodes $P(\sigma > 1) = 0.01$, shrinking each standard deviation toward zero:

\begin{CodeInput}
model = {
    'response': 'Y',
    'fixed': ['1', 'AFF'],
    'random': [{
        'id': 'idarea',
        'model': 'bym',
        'graph': 'output/scotland_connected.adj',
        'scale.model': True,
        'hyper': {
            'theta1': {'prior': 'pc.prec', 'param': [1, 0.01]},
            'theta2': {'prior': 'pc.prec', 'param': [1, 0.01]}
        }
    }]
}

result = pyinla(model=model, family='poisson', data=data, E=data['E'].values,
                control={'compute': {'dic': True, 'waic': True,
                         'config': True, 'return_marginals': True},
                         'fixed': {'prec.intercept': 0.001, 'prec': 0.001}})
\end{CodeInput}

\begin{CodeOutput}
PyINLA fitting time: 0.42 seconds

Fixed effects:
                 mean        sd  0.025quant  0.975quant
(Intercept) -0.264932  0.123754   -0.507552   -0.020795
AFF          4.217436  1.279918    1.662672    6.697430

Hyperparameters:
                                    mean       sd  0.025quant  0.975quant
Prec. idarea (iid)            874.329651  6125.35     7.99932  5891.90199
Prec. idarea (spatial)          4.762095     1.86     2.04940     9.23697

\end{CodeOutput}

The AFF coefficient $\hat{\beta}_1 = 4.22$ indicates that a 10\% increase in
AFF employment is associated with a multiplicative risk increase of
$e^{0.1 \times 4.22} \approx 1.52$. The very high IID precision
($\hat{\tau}_{\text{iid}} \approx 874$, corresponding to
$\hat{\sigma}_{\text{iid}} \approx 0.12$) indicates that nearly all residual
variation is spatially structured rather than independent noise.

The spatial fraction $\phi$ is derived from posterior hyperparameter samples
using \texttt{inla\_hyperpar\_sample}:

\begin{CodeInput}
from pyinla.sampling import inla_hyperpar_sample

hyp_samples = inla_hyperpar_sample(n=100000, result=result, intern=True).values
vv = np.exp(-hyp_samples)  # columns: 1/tau_iid, 1/tau_spatial
phi_samples = vv[:, 1] / vv.sum(axis=1)

print(f"Spatial fraction (phi): {phi_samples.mean():.3f}")
print(f"95% CI: [{np.percentile(phi_samples, 2.5):.3f}, "
      f"{np.percentile(phi_samples, 97.5):.3f}]")
\end{CodeInput}

\begin{CodeOutput}
Spatial fraction (phi): 0.920
95% CI: [0.644, 0.999]
\end{CodeOutput}

The derived $\hat{\phi} = 0.92$ indicates that approximately 92\% of the
random effect variance is spatially structured, consistent with geographic
clustering of risk factors.

\subsubsection{Comparison with MCMC}

To validate PyINLA, we fit the identical scaled BYM model using PyMC with NUTS
(4 chains, 5,000 draws each after 3,000 tuning iterations,
\texttt{target\_accept=0.95}). The PC prior $P(\sigma > 1) = 0.01$ translates
to an Exponential prior on the standard deviation with rate
$\lambda = -\log(0.01) \approx 4.605$:
$\sigma_{\text{spatial}} \sim \text{Exp}(4.605)$,
$\sigma_{\text{iid}} \sim \text{Exp}(4.605)$.

\begin{table}[t]
\centering
\caption{Timing and parameter comparison between PyINLA and MCMC (PyMC/NUTS)
for the scaled BYM disease mapping model.}
\label{tab:scottish-comparison}
\begin{tabular}{lrrrrr}
\hline
& \multicolumn{2}{c}{PyINLA} & \multicolumn{2}{c}{MCMC} & \\
Parameter & Mean & SD & Mean & SD & $|\Delta|$ \\
\hline
\multicolumn{6}{l}{\textit{Timing}} \\
\quad CPU time (s) & \multicolumn{2}{c}{0.42} & \multicolumn{2}{c}{117.2} & 278$\times$ \\
\hline
\multicolumn{6}{l}{\textit{Fixed effects}} \\
\quad Intercept & $-0.265$ & 0.124 & $-0.261$ & 0.122 & 0.004 \\
\quad AFF & 4.217 & 1.280 & 4.170 & 1.269 & 0.047 \\
\hline
\multicolumn{6}{l}{\textit{Hyperparameters (SD scale)}} \\
\quad $\sigma_{\text{spatial}}$ & 0.450 & --- & 0.474 & 0.091 & 0.024 \\
\quad $\sigma_{\text{iid}}$ & 0.120 & --- & 0.119 & 0.090 & 0.001 \\
\hline
\multicolumn{6}{l}{\textit{Derived}} \\
\quad $\phi$ & 0.920 & --- & 0.907 & --- & 0.013 \\
\hline
\end{tabular}
\end{table}

Table~\ref{tab:scottish-comparison} shows that PyINLA completes model fitting
in 0.42~seconds compared to 117.2~seconds for MCMC, a speedup of approximately
$278\times$. Fixed effects agree closely (AFF coefficient difference is only
0.04 posterior standard deviations), and the Pearson correlation across all 56
district-level relative risks is $r = 0.9999$, demonstrating excellent
agreement between INLA and MCMC with identical priors on both precision
hyperparameters.

\subsubsection{Disease risk mapping}

\begin{figure}[t!]
\centering
\includegraphics[width=0.85\textwidth]{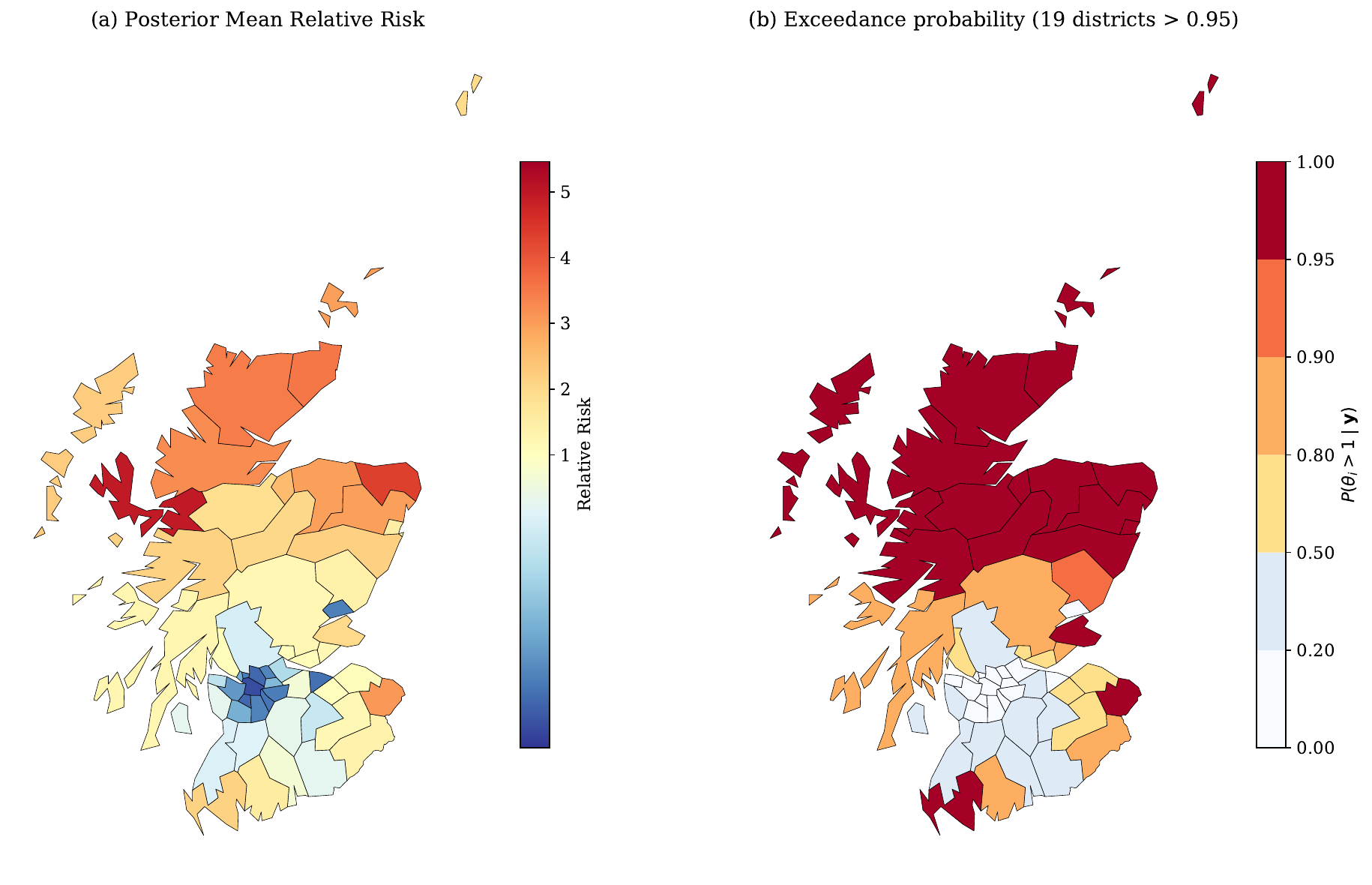}
\caption{Left: posterior mean relative risk for lip cancer across 56 Scottish
districts (scaled BYM model). Colors diverge from RR~$= 1$, with red indicating
elevated risk and blue indicating reduced risk. Right: exceedance probability
$P(\theta_i > 1 \mid \boldsymbol{y})$. The spatial pattern reflects both the
AFF covariate effect and residual spatial correlation.}
\label{fig:scottish-map}
\end{figure}

Figure~\ref{fig:scottish-map} displays choropleth maps of the posterior mean
relative risk and exceedance probability across the 56 Scottish districts.
Red shading indicates elevated risk (RR~$> 1$) and blue indicates reduced risk.
High-risk areas cluster in northern agricultural regions, particularly
Skye-Lochalsh, Banff-Buchan, and Caithness, where AFF employment is highest.

A key advantage of PyINLA is direct access to exceedance probabilities. For
each district, we compute $P(\theta_i > 1 \mid \boldsymbol{y})$, the posterior
probability that the true relative risk exceeds 1:

\begin{CodeInput}
import pyinla as pinla

for i in range(len(data)):
    exceedance = 1.0 - pinla.pmarginal(1.0, result.marginals_fitted_values[i])
    data.loc[i, 'exceedance_prob'] = exceedance
\end{CodeInput}

Of the 56 districts, 31 have exceedance probability $> 0.95$, providing
strong evidence of elevated risk.

\subsubsection{Model fit}

The scaled BYM model yields DIC $= 297.63$ and WAIC $= 294.64$. The spatial
smoothing substantially reduces the noise in crude SIRs: for example,
Skye-Lochalsh has a crude SIR of 6.43 but a smoothed relative risk of 4.66,
borrowing strength from neighboring districts. \\

PyINLA and MCMC produce consistent estimates, with district-level relative
risks correlated at $r = 0.9999$ and the AFF coefficient differing by only
0.04 posterior standard deviations. PyINLA completes the fit in 0.42~seconds
compared to 117~seconds for MCMC, a $278\times$ speedup. The derived spatial
fraction $\hat{\phi} = 0.92$ confirms that 92\% of residual variation is
spatially structured, consistent with the geographic clustering of outdoor
work exposure across Scotland.

\paragraph{Replication materials.}
The complete replication notebook is available as supplementary material
(\texttt{scottish.py}).

\subsection{Geostatistical modeling: regional temperature interpolation}
\label{sec:ex-spde}

This example demonstrates continuous spatial prediction using the SPDE 
(stochastic partial differential equation) approach \citep{lindgren2011explicit}, 
applied to January 2024 temperature data from 55 weather stations across 
Lebanon, Syria, Jordan, and Saudi Arabia. Data were obtained from the 
Global Historical Climatology Network \citep{menne2012overview}.

\subsubsection{Model specification}

Let $T(s_i)$ denote mean temperature at station location $s_i$. We specify:
\begin{equation}\label{eq:spde-model}
T(s_i) = \beta_0 + \beta_1 \, z_i + u(s_i) + \varepsilon_i,
\end{equation}
where $\beta_0$ is the intercept (baseline temperature at sea level), 
$z_i = \text{elevation}_i / 1000$ is elevation in kilometers, 
$u(s)$ is a Gaussian random field with Mat\'ern covariance approximated 
via the SPDE representation, and $\varepsilon_i \sim \mathcal{N}(0, \sigma^2_\varepsilon)$ 
is measurement error. We expect $\beta_1 \approx -6$ based on the standard 
atmospheric lapse rate of approximately $6^\circ$C per 1000\,m.

The spatial field $u(s)$ is assigned penalized complexity (PC) priors 
\citep{simpson2017penalising} on the practical range $\rho$ and marginal 
standard deviation $\sigma_u$:
\begin{equation}
\Pr(\rho < 30\text{ km}) = 0.05, \qquad \Pr(\sigma_u > 1^\circ\text{C}) = 0.05.
\end{equation}

\subsubsection{Implementation}

We load the temperature data and project coordinates to UTM Zone 40 
(units in kilometers):

\begin{CodeInput}
import numpy as np
import pandas as pd
import geopandas as gpd
from scipy import sparse
from shapely.geometry import Point
from pyinla import pyinla
from pyinla.fmesher import (fm_mesh_2d, fm_hexagon_lattice,
    spde2_pcmatern, spde_make_A, spde_grid_projector)

# Coordinate reference system (UTM Zone 40, units in km)
CRS_KM = "+proj=utm +zone=40 +units=km +no_defs"

# Load temperature data
tavg = pd.read_csv("tavg_month.csv")

# Project to km-based CRS
geometry = [Point(lon, lat) for lon, lat in 
    zip(tavg["longitude"], tavg["latitude"])]
tavg_gdf = gpd.GeoDataFrame(tavg, geometry=geometry, crs="EPSG:4326")
tavg_km = tavg_gdf.to_crs(CRS_KM)
coords_km = np.column_stack([tavg_km.geometry.x, tavg_km.geometry.y])

print(f"Stations: {len(tavg)}")
print(f"Temperature range: {tavg['X202401'].min():.1f} to "
      f"{tavg['X202401'].max():.1f} C")
\end{CodeInput}

\begin{CodeOutput}
Stations: 55
Temperature range: 9.0 to 26.1 C
\end{CodeOutput}

We construct a triangular mesh using a hexagonal lattice over the buffered 
study region. The mesh extends 500\,km beyond the data to avoid boundary 
effects:

\begin{CodeInput}
# Create hexagonal lattice within buffered boundary
boundary_buffered = boundary.buffer(100)
boundary_coords = np.array(boundary_buffered.exterior.coords)
hexpoints = fm_hexagon_lattice(boundary_coords, edge_len=50)

# Build triangular mesh
mesh = fm_mesh_2d(loc=hexpoints, offset=500, max_edge=200)
print(f"Mesh: {mesh.n} vertices, {mesh.n_triangle} triangles")
\end{CodeInput}

\begin{CodeOutput}
Mesh: 1792 vertices, 3499 triangles
\end{CodeOutput}

\begin{figure}[t!]
\centering
\includegraphics[width=0.7\textwidth]{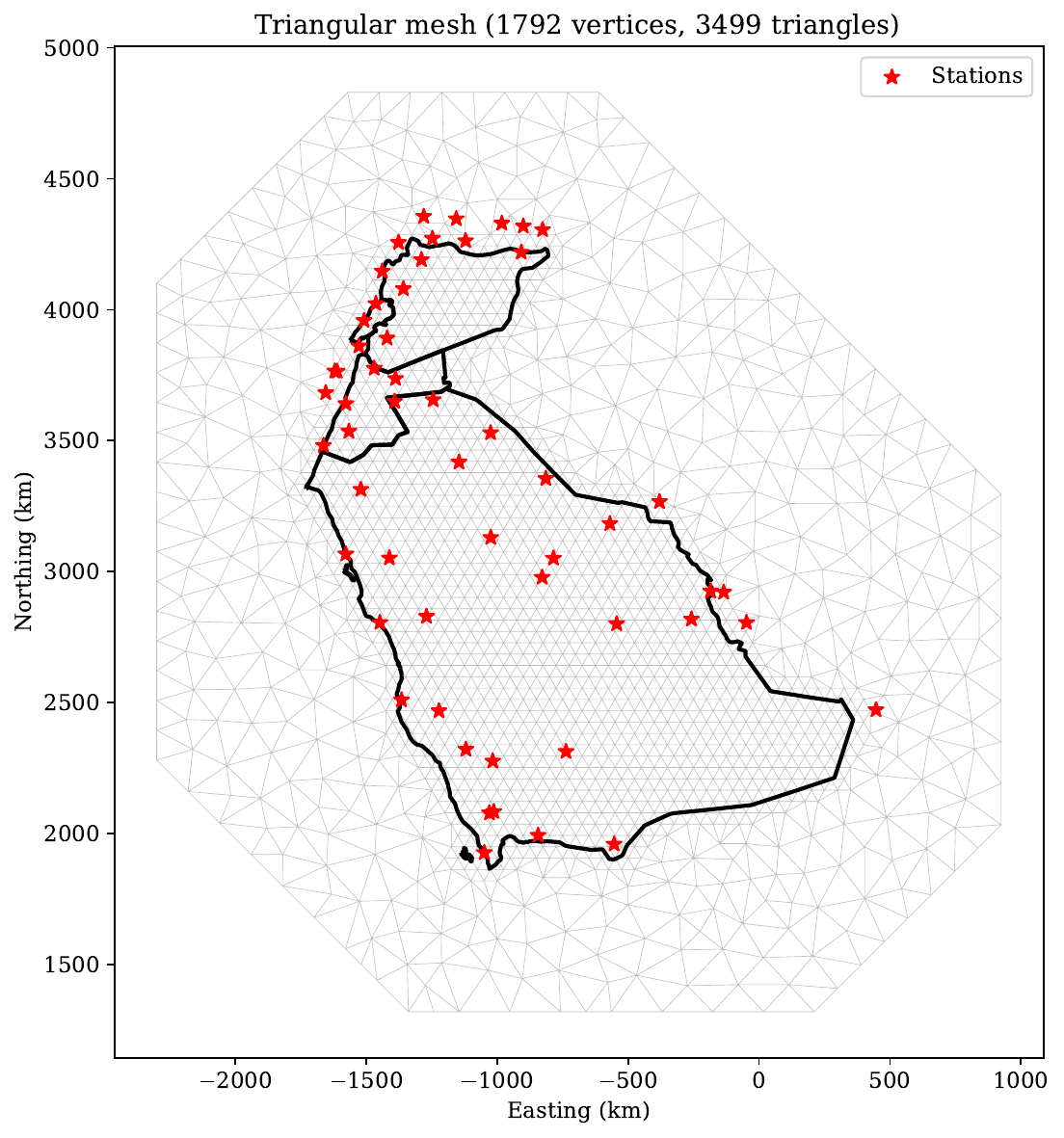}
\caption{Triangular mesh for SPDE approximation of the Mat\'ern field. 
The mesh contains 1792 vertices and 3499 triangles, with finer resolution 
near the study region boundary and coarser triangles in the outer extension. 
Red stars indicate weather station locations.}
\label{fig:spde-mesh}
\end{figure}

Figure~\ref{fig:spde-mesh} shows the resulting mesh along with the 55 
weather station locations.

The SPDE model with PC priors and observation projection matrix are defined as:

\begin{CodeInput}
# Define SPDE model with PC priors
spde = spde2_pcmatern(
    mesh=mesh,
    prior_range=(30, 0.05),   # P(range < 30 km) = 0.05
    prior_sigma=(1, 0.05)     # P(sigma > 1 C) = 0.05
)

# Create projection matrix from mesh to observation locations
A = spde_make_A(mesh=mesh, loc=coords_km)
\end{CodeInput}

Model fitting uses a dictionary-based specification:

\begin{CodeInput}
# Prepare data
n_obs = len(tavg)
dataf = pd.DataFrame({
    'X202401': tavg['X202401'].values,
    'elevation_km': tavg['elevation'].values / 1000,
    'spatial': np.arange(n_obs)
})

# Define model: temperature ~ intercept + elevation + spatial effect
model = {
    'response': 'X202401',
    'fixed': ['1', 'elevation_km'],
    'random': [{'id': 'spatial', 'model': spde, 'A.local': A}]
}

# Fit model
result = pyinla(model=model, family='gaussian', data=dataf)
print(result.summary_fixed[['mean', 'sd', '0.025quant', '0.975quant']])
\end{CodeInput}

\begin{CodeOutput}
                 mean     sd  0.025quant  0.975quant
(Intercept)     20.12   3.87       12.10       28.14
elevation_km    -5.74   0.37       -6.47       -5.01
\end{CodeOutput}

The estimated lapse rate of $-5.74^\circ$C per kilometer 
(95\% CI: $[-6.47, -5.01]$) is consistent with the atmospheric expectation. 
The intercept of $20.12^\circ$C represents the predicted January temperature 
at sea level; its wider uncertainty (SD $= 3.87$) reflects the spatial 
correlation structure absorbing much of the intercept variability.

\subsubsection{Spatial prediction}

We generate predictions on a regular 10\,km grid (230 $\times$ 280 points), 
incorporating elevation from ETOPO 2022 digital elevation data \citep{noaa2022etopo}:

\begin{CodeInput}
# Create prediction grid projector
projGrid = spde_grid_projector(
    mesh=mesh, xlim=(-1800, 500), ylim=(1700, 4500), dims=(230, 280))

# Prediction data: NA responses, grid elevation from ETOPO 2022
n_pred = len(projGrid.lattice)
preddf = pd.DataFrame({
    'X202401': np.full(n_pred, np.nan),
    'elevation_km': etopoll / 1000,
    'spatial': np.arange(n_obs, n_obs + n_pred)
})
data_combined = pd.concat([dataf, preddf], ignore_index=True)

# Combine observation and prediction projection matrices
A_combined = sparse.vstack([A, projGrid.proj])

# Same model structure with the combined projection matrix
model_pred = {
    'response': 'X202401',
    'fixed': ['1', 'elevation_km'],
    'random': [{'id': 'spatial', 'model': spde, 'A.local': A_combined}]
}
result_pred = pyinla(model=model_pred, family='gaussian', data=data_combined)

# Extract predictions (rows after the observations)
i_pred = range(n_obs, n_obs + n_pred)
pred_mean = result_pred.summary_fitted_values.iloc[i_pred]['mean'].values
pred_sd = result_pred.summary_fitted_values.iloc[i_pred]['sd'].values
\end{CodeInput}

\begin{CodeOutput}
Temperature range: -2.5 to 27.7 C
Uncertainty range: 0.38 to 1.46 C
\end{CodeOutput}

\begin{figure}[t!]
\centering
\includegraphics[width=\textwidth]{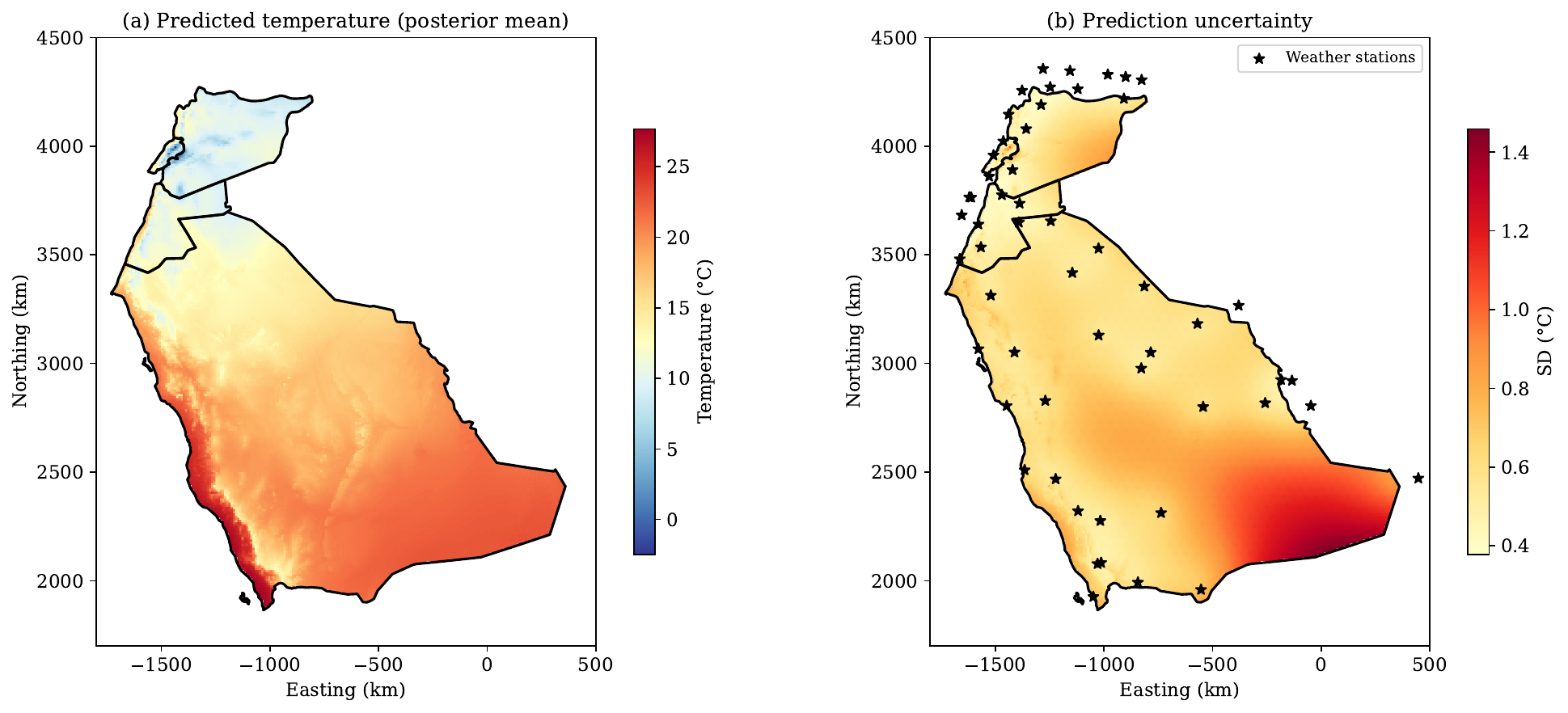}
\caption{Spatial temperature predictions for January 2024. Left: posterior 
mean temperature showing elevation-driven gradients from warm lowlands 
(25--28$^\circ$C) to cool highlands (below 10$^\circ$C). Right: prediction 
uncertainty (posterior standard deviation), lowest near weather stations 
(black stars) and highest in data-sparse regions.}
\label{fig:spde-results}
\end{figure}

Figure~\ref{fig:spde-results} displays the predicted temperature surface 
and associated uncertainty. Prediction uncertainty is lowest near observation 
locations and increases in data-sparse regions, particularly in interior 
Saudi Arabia where the nearest stations are several hundred kilometers away.

\begin{figure}[t!]
\centering
\includegraphics[width=0.5\textwidth]{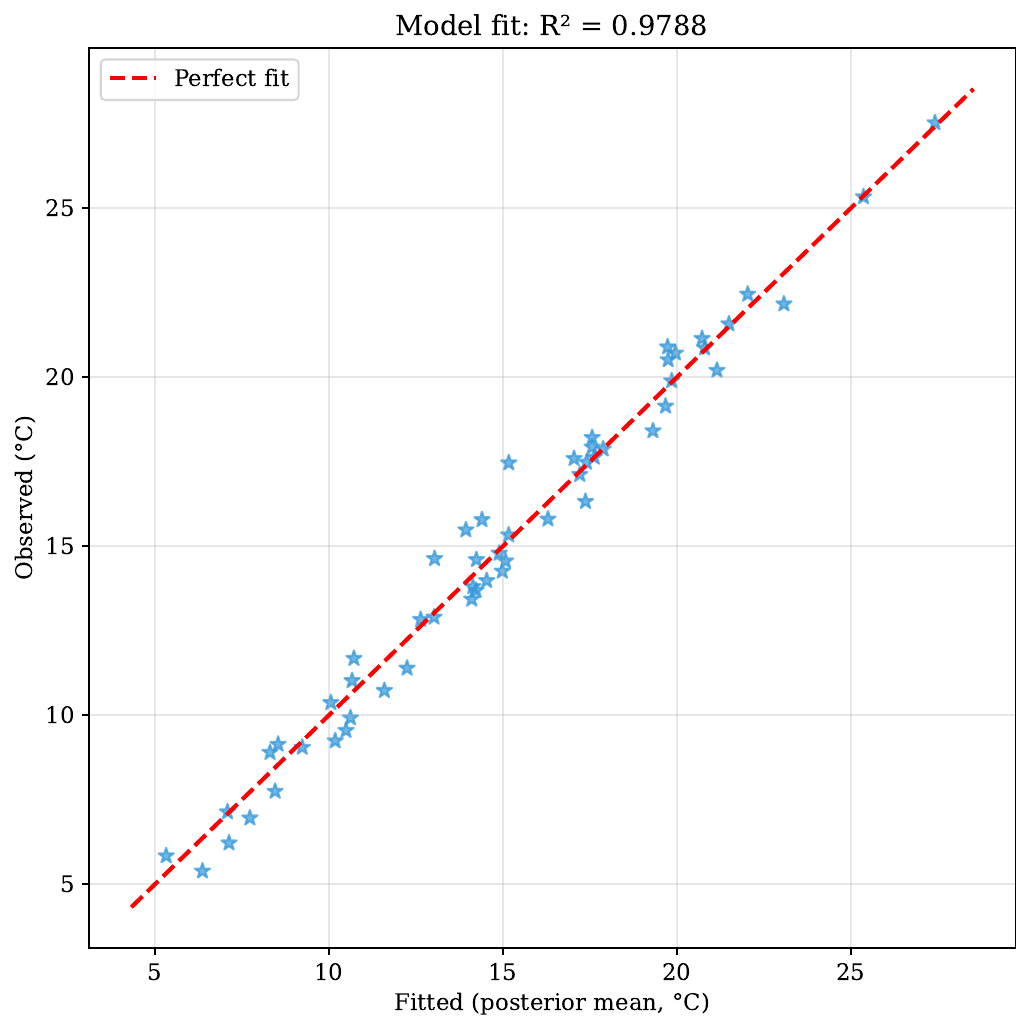}
\caption{Observed versus fitted temperatures showing excellent model fit 
($R^2 = 0.979$). The dashed line indicates perfect agreement.}
\label{fig:spde-fitted}
\end{figure}

The model achieves $R^2 = 0.979$ for fitted versus observed temperatures 
(Figure~\ref{fig:spde-fitted}), with predicted values ranging from 
$-2.5^\circ$C in mountainous areas to $27.7^\circ$C in low-elevation 
desert regions.

\paragraph{Replication materials.}
The complete replication script, including data acquisition from GHCN and 
all intermediate processing steps, is available as supplementary material 
(\texttt{spde.py}).

\section{Discussion and conclusion} \label{sec:conclusion}

This paper introduced \textsf{PyINLA}, a native \proglang{Python} package that brings INLA-based Bayesian inference to the \proglang{Python} ecosystem. By providing direct access to the compiled INLA computational engine through a Pythonic API, \textsf{PyINLA} eliminates the friction of cross-language workflows while preserving the computational efficiency and statistical accuracy that have made INLA a standard tool for latent Gaussian models.

The examples in Section~\ref{sec:examples} demonstrated \textsf{PyINLA} across a range of model structures, from exchangeable random effects for sports prediction to spatially structured effects for disease mapping, continuous spatial fields for geostatistical interpolation, and temporal components for time series forecasting. Across these applications, posterior estimates closely 
matched MCMC benchmarks while achieving computational speedups of 100--200$\times$, with deterministic outputs that simplify testing and deployment.

Several limitations should be acknowledged. \textsf{PyINLA} is restricted to latent Gaussian models; applications requiring non-Gaussian latent structure or likelihoods outside INLA's supported families require general-purpose tools. Applications requiring direct access to the full joint posterior may also benefit from MCMC methods, although \textsf{PyINLA} provides posterior sampling for derived quantities.

Development continues along several axes, including extended model support, GPU acceleration via the sTiles library \citep{fattah2025stiles, fattah2025gpu}, deeper integration with \proglang{Python} machine learning workflows, and expanded documentation at \url{https://pyinla.org}. Future releases will also allow users to define custom latent models and likelihood functions, extending \textsf{PyINLA} beyond the built-in components.

\textsf{PyINLA} makes INLA-based Bayesian inference a native part of the \proglang{Python} scientific computing ecosystem. By providing direct access to fast, deterministic inference for latent Gaussian models, \textsf{PyINLA} opens the door to tighter integration with modern machine learning and AI workflows, where principled uncertainty quantification is increasingly valued. We anticipate that native \proglang{Python} availability will expand the reach of INLA methodology to new application domains and 
larger-scale problems.

\bibliographystyle{plainnat}
\bibliography{mybib}

\newpage
\appendix

\section{Marginal posterior utility functions}\label{app:marginals}

After model fitting, \textsf{PyINLA} returns marginal posterior distributions
$p(\theta_i \mid \mathbf{y})$ for each model component (fixed effects,
hyperparameters, random effects, and linear predictor values). Each marginal
is stored as an $(n \times 2)$ \pkg{NumPy} array of $(x, \text{density})$ pairs.
The following utility functions allow users to work with these marginals
as continuous distributions.

\subsection{Density evaluation}

The function \code{dmarginal(x, marginal)} evaluates the posterior density
at one or more points using monotone piecewise cubic Hermite interpolation (PCHIP):
\begin{equation}
d(x) = p(\theta = x \mid \mathbf{y}).
\end{equation}
For points outside the support, the function returns zero.

\subsection{Cumulative distribution}

The function \code{pmarginal(x, marginal)} computes the cumulative probability:
\begin{equation}
F(x) = P(\theta \leq x \mid \mathbf{y}) = \int_{-\infty}^{x} p(\theta \mid \mathbf{y}) \, d\theta.
\end{equation}
This is useful for posterior hypothesis assessment. For example, for a
regression coefficient $\beta$, the quantity $1 - F(0)$ gives the posterior
probability that $\beta$ is positive.

\subsection{Quantile function}

The function \code{qmarginal(p, marginal)} returns the inverse CDF:
\begin{equation}
Q(p) = F^{-1}(p) = \inf\{x : F(x) \geq p\}.
\end{equation}
Equal-tailed credible intervals are obtained as $[Q(\alpha/2),\; Q(1-\alpha/2)]$.

\subsection{Transformation}

The function \code{tmarginal(g, marginal)} transforms a marginal through a
monotonic function $g$, returning the marginal of $Y = g(\theta)$ via the
change-of-variables formula:
\begin{equation}
p_Y(y) = p_\theta(g^{-1}(y)) \cdot \left|\frac{d}{dy}g^{-1}(y)\right|.
\end{equation}
A common use is converting log-scale parameters to the natural scale
(e.g., log-precision to standard deviation via $g(\theta) = \exp(-\theta/2)$),
or computing odds ratios from log-odds via $g(\theta) = \exp(\theta)$.

\subsection{Expected value of a function}

The function \code{emarginal(g, marginal)} computes:
\begin{equation}
\mathbb{E}[g(\theta) \mid \mathbf{y}] = \int g(\theta) \cdot p(\theta \mid \mathbf{y}) \, d\theta,
\end{equation}
using numerical quadrature on the discrete marginal.

\subsection{Highest posterior density interval}

The function \code{hpdmarginal(level, marginal)} returns the shortest
interval containing probability mass $1-\alpha$:
\begin{equation}
\text{HPD}_{1-\alpha} = \{x : p(x \mid \mathbf{y}) \geq k_\alpha\},
\end{equation}
where $k_\alpha$ is chosen such that $P(\theta \in \text{HPD}_{1-\alpha}) = 1 - \alpha$.
For skewed posteriors, HPD intervals are shorter than equal-tailed intervals
at the same coverage level.

\subsection{Summary statistics}

The function \code{zmarginal(marginal)} returns a dictionary containing the
posterior mean, standard deviation, median, and $2.5\%$/$97.5\%$ quantiles,
computed from the discrete marginal representation.

\subsection{Random sampling and mode}

The function \code{rmarginal(n, marginal)} generates $n$ random samples
using inverse transform sampling: $X = F^{-1}(U)$ where $U \sim \text{Uniform}(0,1)$.
The function \code{mmarginal(marginal)} returns the posterior mode (MAP estimate):
\begin{equation}
\hat{\theta}_{\text{MAP}} = \arg\max_\theta \; p(\theta \mid \mathbf{y}).
\end{equation}

\subsection{Usage example}

The following illustrates a typical workflow after fitting a model:

\begin{CodeChunk}
\begin{CodeInput}
import pyinla
import numpy as np

# Access the marginal for a fixed effect
marg = result.marginals_fixed["x"]

# Summary statistics
summary = pyinla.zmarginal(marg)

# Posterior probability that the coefficient is positive
prob_positive = 1.0 - pyinla.pmarginal(0.0, marg)

# 95% highest posterior density interval
hpd95 = pyinla.hpdmarginal(0.95, marg)

# Transform from log-odds to odds ratio
marg_or = pyinla.tmarginal(np.exp, marg)

# Expected value of a derived quantity
e_or = pyinla.emarginal(np.exp, marg)
\end{CodeInput}
\end{CodeChunk}

\end{document}